# Power system investment optimization to identify carbon neutrality scenarios for Italy

Alice Di Bella[1,2,3*], Federico Canti[1,2], Matteo Giacomo Prina[1], Valeria Casalicchio[1], Giampaolo Manzolini[2], Wolfram Sparber[1]

[1]*Institute for Renewable Energy, EURAC Research, Viale Druso 1, I-39100 Bolzano, Italy*
[2]*Dipartimento di energia, Politecnico di Milano, Via Lambruschini, 4, 20156 Milano (MI), Italy*
[3]*RFF-CMCC European Institute on Economics and the Environment (EIEE), Centro Euro-Mediterraneo sui Cambiamenti Climatici, Italy*
[*]*E-mail : alicedb7@gmail.com*

## Abstract

*In 2021 the European Commission has adopted the Fit-for-55 policy package, legally binding European countries to reduce their $CO_2$ emissions by 55% with respect to 1990, a first step to achieve carbon neutrality by 2050. In this context it is crucial to help national policymakers to choose the most appropriate technologies to achieve these goals and energy system modelling can be a valuable tool.* This article presents a model of the Italian power system realized employing the open energy modelling framework, Oemof. A Linear Programming Optimization is implemented to evaluate how to minimize system costs at decreasing $CO_2$ emissions *in 2030*. The developed tool is applied to evaluate different research questions: i) pathway towards full decarbonization and power self-sufficiency of the electricity sector in Italy, ii) relevance of flexibility assets in power grids: li-ion batteries, hydrogen storage and transmission lines reinforcement. A 55% $CO_2$ emissions reduction for the actual Italian power sector can be achieved through an increase of 30% of the total annual system cost. Full decarbonization can be reached with four times today's annual costs, which could be lowered with sector coupling or considering more technologies.

## 1 Introduction

The consequences of the increase in GHG emissions are getting to a tipping point and human activities are the flagship accountable for climate change. According to the *Intergovernmental Panel on Climate Change* (IPCC) 2021 report [1], it is only possible to avoid warming of 1.5 °C or 2 °C if massive and immediate cuts in greenhouse gas emissions are made. Nevertheless, with the current measures, the World will experience a further temperature rise of 3 °C [2], far beyond the *Paris Agreement* objective of limiting the increase to 1.5 Celsius degrees [3].

*The international community has joint efforts to fight this global challenge, promoting cooperation between countries and international climate policies to push a shift to a more sustainable global energy system. In this context, energy system models can help decision makers to shape efficiently their transition strategies*: they can be crucial in the optimization of ambitious but necessary plans for future energy supply and demand in low-carbon systems and for increasing social acceptance of innovative policies and subsidies. The energy sector accounts for more than 73% of the global $CO_2$ equivalent emissions [4], thus it is paramount to act in this area to abate the human impact on the environment. In this sector, the electricity field is crucial for $CO_2$ emission reduction, thanks to mature and well-known green technologies and the extensive know-how already delineated.

This work models the Italian power system in 2030, divided into 7 market zones and considering the whole year of electricity generation with an hourly resolution. An expansion capacity optimization, starting from the existing technologies helps evaluating the most convenient investment strategy for different degrees of decarbonization and highlighting crucial technologies which need to be enforced. This study has the purpose to highlight the most cost-effective and solid strategy to achieve the European decarbonization target of a 62% reduction of $CO_2$ emissions from the power sector [5]. Furthermore, reaching carbon neutrality is the leitmotif of the present paper, taking into account generation technologies such as onshore wind, rooftop and utility scale photovoltaics, in line with the Italian Integrated National Energy and Climate Plan [6]. The model can invest in storage options like batteries and hydrogen, with the possibility to enhance the transmission grid and blending hydrogen into the methane pipelines (Power-to-Gas). The GitHub repository [7] presents the code and inputs for the Oemof Italy model.

Numerous papers in literature offer similar analyses, with different methodologies, resolutions and details. A similar approach to the energy policy, underlining pros and cons of the adopted transition protocols, is carried out by Mearns and Sornette [8] for the Swiss power system, analysing two months of generation in January and July 2050. The first interesting study in this literature review was carried out by Prina et al. [9], coupling the Oemof framework with a multi-objective optimization (MOO), using an evolutionary algorithm to simulate the Italian energy system with six nodes without reaching the complete decarbonization of the energy system. Another study on the Italian energy system is conducted by Bellocchi et al. [10], in this case with a single node. Hilpert et al. [11] applied linear programming (LP) for the Jordanian electricity system with a single-node approach and an hourly resolution. Herc et al. [12] employed the modelling tool H2RES to perform a linear optimization of the Croatian energy system, with hourly time steps and also considering sector coupling. Louis et al. [13] carried out a study of the European power system considering a multi-node scheme. They used a low temporal resolution (12 typical days), introducing some approximations that need to be quantified. In the work performed by Frysztacki et al. [14] an analysis of the European electricity system and the evaluation of the effects of high wind and solar penetration are inspected with a three-hour timesteps. Finally, Victoria et al. [15] conducted an insightful study of the European energy framework. Their research represents a multi-node network of 30 different nodes and considers an hourly temporal resolution.

In this work, one of the key results is the development of a comprehensive model of the Italian power system that through linear optimization is able to analyze the pathway towards full decarbonization and power self-sufficiency, with high spatial and temporal resolution. The regional disaggregation allows to evaluate where is more cost-efficient to implement new generation and storage technologies. A sensitivity analysis on Power-to-Gas technology has been conducted as there is a significant amount of uncertainty surrounding the proportions of hydrogen that can be safely and efficiently injected into the gas grid.

This work is organized as follows: in Section 2 an brief overview of the Oemof framework is presented. The third section proposes a deep investigation of the Italian power sector case, specifying input data. In Section 4, the outcomes of the expansion capacity optimization and sensitivity analyzes are delineated. To conclude, Section 5 summarizes the primary outcomes of the performed research, highlighting the remarkable results and listing the possibilities for further development of the conducted study.

# 2 Methodology and Data

## 2.1 Oemof model

*Open Energy MOdelling Framework* (Oemof) is an open-source energy modelling tool developed by Reiner Lemoine Institut and the Center for Sustainable Energy Systems Flensburg [16]. It is made up by a collection of python libraries, each one with a specific task in the modeling process: for example the library oemof-network sets up the energy connections and oemof-solph builds the LP/MILP energy system model. The framework requires a solver; in this case, Gurobi was the one used. Oemof framework needs the definition of the following input:

- *Energy demand*. For each time step, Oemof requests a value for peak demand in each node and a time series of values from 0 to 1 for each time slice, describing the energy load distribution. This work implements an hourly temporal resolution and a time horizon of one year; thus the model relies on 8760 inputs for each node and a peak value in MW. A sensitivity analysis to evaluate the importance of a granular time resolution for system with a large penetration of RES is offered in Appendix C.

- *Variable RES plants*. The availability distributions for unpredictable renewable energy sources can be specified through a series of values between 0 and 1 (0 being no availability, 1 producing at nominal capacity). Since the capacity of RES power plants can be increased in the investment scenarios, a maximum potential is provided, divided into different regions.

- *Power plants and storage*. For each node and each type of technology, installed power plants capacity and installed storage energy are provided in MW and MWh. For the latter, the input and output capacities are specified too. For fossil fuels power plants, which are a *Transformer* component, it is crucial to specify efficiency and the fuel used. In contrast, the nominal power corresponds to the net possible power output for renewable plants. Unlimited maximum potential for batteries and hydrogen tanks is assumed since these technologies do not rely directly on natural sources. $H_2$ storage requires additionally the definition of two transformers to model the conversion of electricity to hydrogen and vice versa: electrolyzers and fuel cells.

- *Transmission lines*. Oemof allows the definition of powerlines linking different regions, taking exchange capacity as input. In this work, they are specified in MW and a transport efficiency is provided. In the investment optimization, powerlines can be expanded, so a capital cost in €/MW is introduced.

- *Technologies investment and operation costs*. O&M costs are provided in €/MWh; capital costs are in €/MW and €/MWh for storage and they already include fixed O&M costs. There are also costs related to the acquirement of fuels, in this case, natural gas and coal.

- *Emission factors for fossil fuel sources*. $CO_2$ emissions coefficients are communicated for polluting energy sources, expressed in €/$MWh_{th}$.

A Single-Objective Optimization is adopted to identify the optimal scenario of the Italian system. *For a comprehensive discussion about equations and constraints employed to optimize the model, please refer to the Appendix A.*

## 2.2 Italian case study

The model of the Italian power sector needs to be validated before being employed to test European policies. The Italian Energy system in 2019 is identified as a validation case study since it does not show the impact of the COVID-19 pandemic, which could cause anomalies due to the atypical energy and electricity usage trend. The Italian electricity framework is divided into 7 market zones consistent with the Italian electricity scenario in 2030 and connected through high voltage powerlines [17,18], as reported in Figure 1.

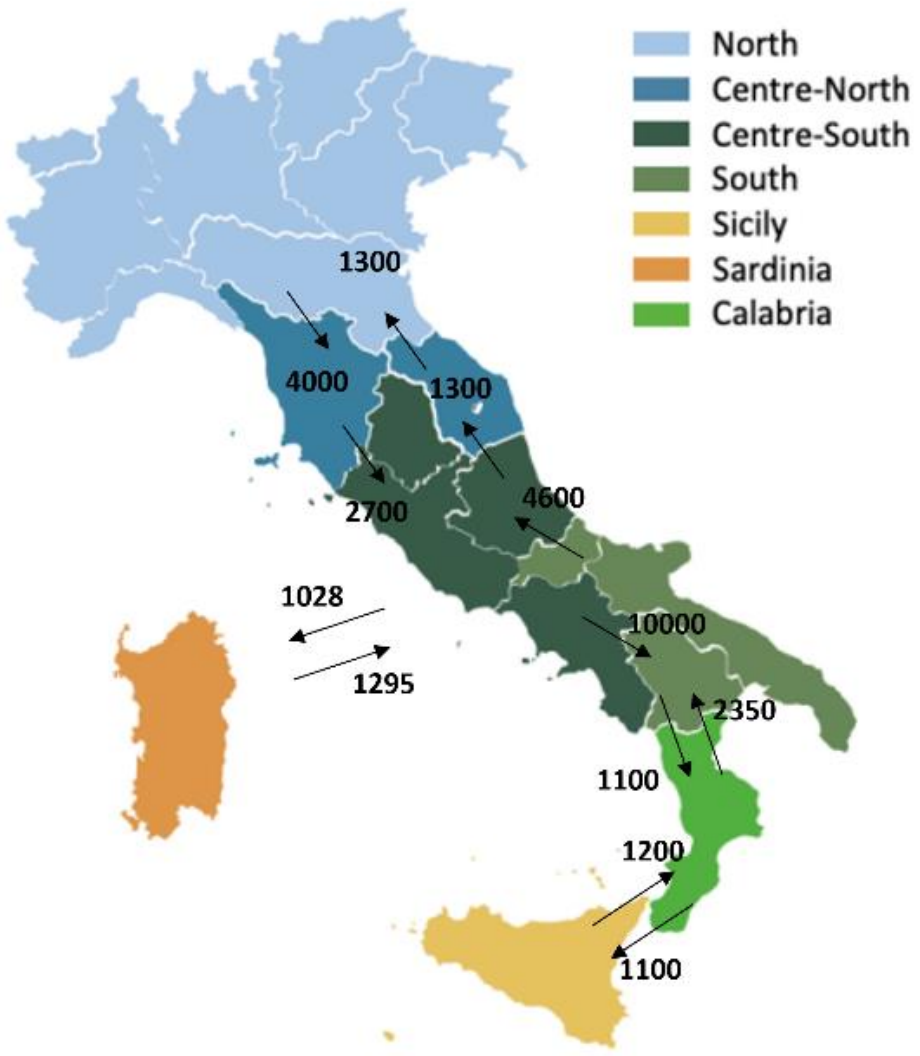

*Figure 1 - Italian electricity market zones in 2021 and powerlines capacity in MW*

Input data for electric demand [19], power plants [20–24], efficiencies [9,25,26], transmission lines [17] and costs [25,27,28] have been taken from different databases. For model validation, data are taken for the 2019 scenario to compare it with the historical electricity production from Terna [29]. An important note must be made regarding the input data for electric demand since current data are aggregated for different market zones. They are, in fact, modified according to the changes and all formulas and calculations are listed in Appendix B.

The operational scheme of the Italian electricity system has been modelled in Oemof as in Figure 2: renewables sources directly produce electricity, while commodities like fossil fuels, reservoir hydro and import pass through a transformer, with their related efficiency. Import refers to the net electricity imported from abroad and has been designed as a generation source with its proper emission factor. Full decarbonization scenarios do not include import in generation, since it has an emission factor that considers the fossil sources employed in other countries to produce electricity, thus also reaching power self-sufficiency for Italy. According to Terna [30] the major importer for Italy is France, so we assumed import as a source located in the North, but transmission lines have the possibility to transport electricity in other regions. PHS is the only source of storage for the Italian power system in 2019.

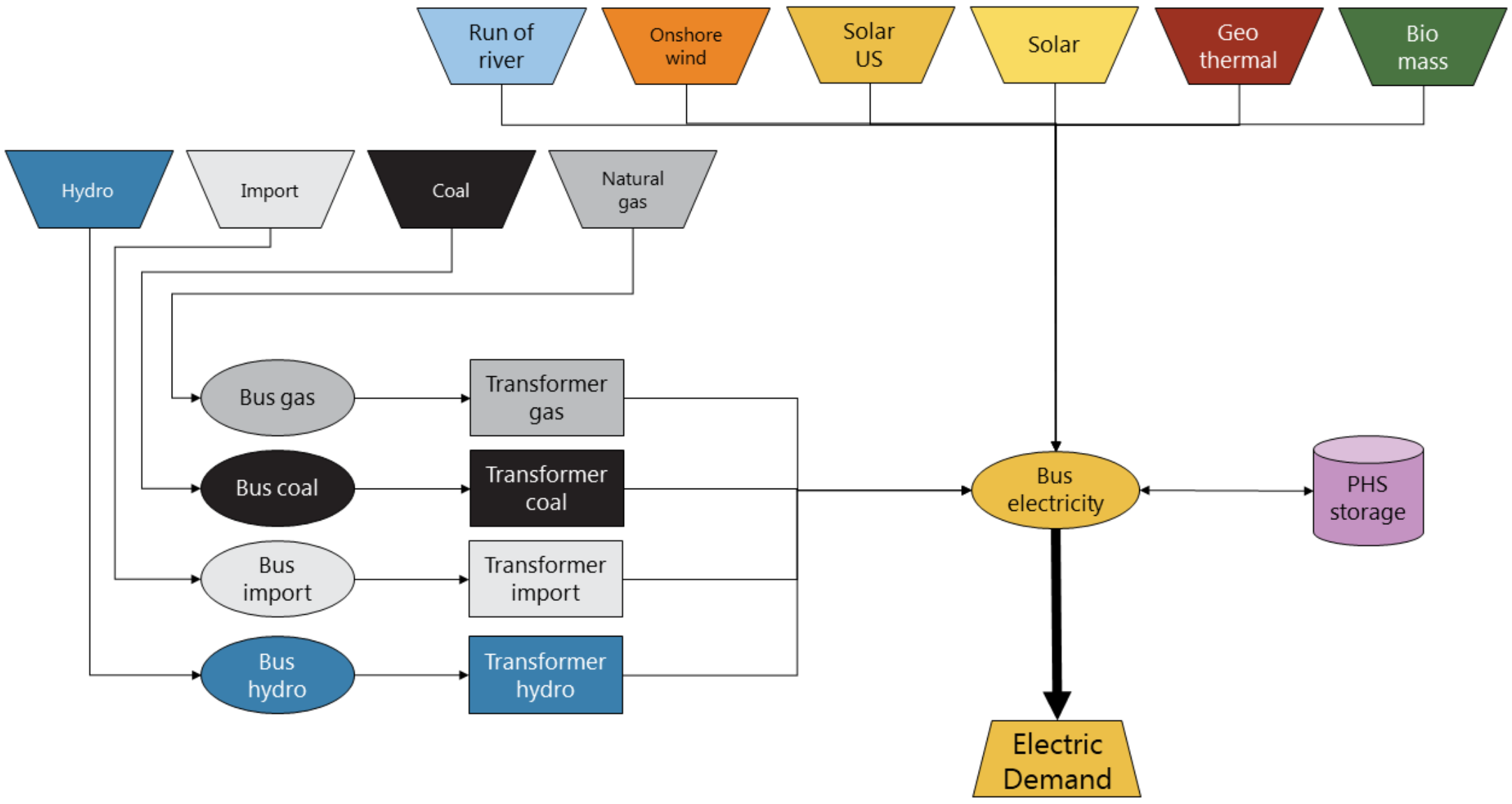

*Figure 2 – Oemof operational scheme for the Italian power sector in 2019*

To check for model validity and reliability, results for the operation optimization are compared with production data from the transmission system operator (TSO) Terna [31] (cf. Figure 3). In Table 1 a breakdown of the energy generation for the different technologies is shown to have a numerical comparison.

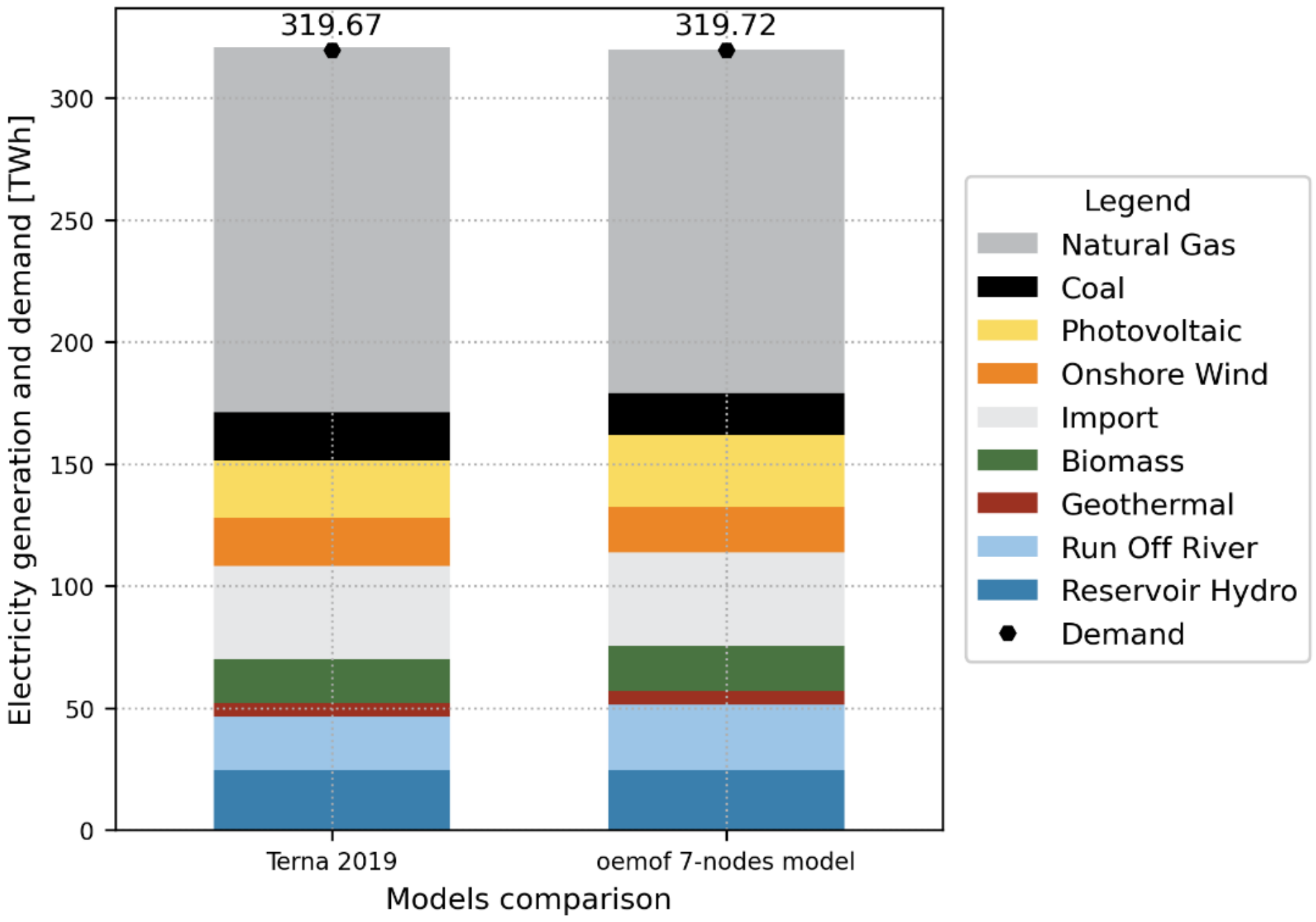

*Figure 3 – Model validation and comparison with Terna electricity mix*

| *Technology* | *Terna 2019 [TWh]* | *7-nodes model [TWh]* |
|---|---|---|
| Natural Gas | 149.6 | 140.7 |
| Coal | 19.65 | 17.24 |
| Photovoltaic | 23.32 | 29.44 |
| On-shore wind | 20.03 | 18.78 |
| Import | 38.14 | 38.14 |
| Biomass | 17.97 | 18.75 |
| Geothermal | 5.690 | 5.554 |
| Run Off River | 21.75 | 26.67 |
| Reservoir Hydro | 24.61 | 24.61 |

*Table 1 - Model validation: values and comparison with Terna*

Figure 3 and Table 2 illustrate that the model results are consistent with actual data. Maximum errors for any generation technology stay below 7%, besides photovoltaic and run-off river productions, which have a discrepancy of +26% and +22%, respectively. In this paper, model data on timeseries and equivalent hours reflect an average over five years, taken from [9], thus they are not exactly the same as the actual sun and wind availability in the year 2019 [32], explaining the differences. The model has also been validated from a $CO_2$ emissions point of view since the emissions output stays within the band gap found when analyzing various references [33–38].

Once the model has been validated, it is possible to introduce the expansion capacity optimization to analyze the electricity mix in 2030 for different $CO_2$ abatements.

The Italian electricity scheme is therefore modified as it is shown in Figure 5: coal power plants have been phased out (according to [6]), storage technologies such as Li-ion batteries and hydrogen tanks have been considered as decision variables and the Power-to-Gas technology has been introduced as an option.

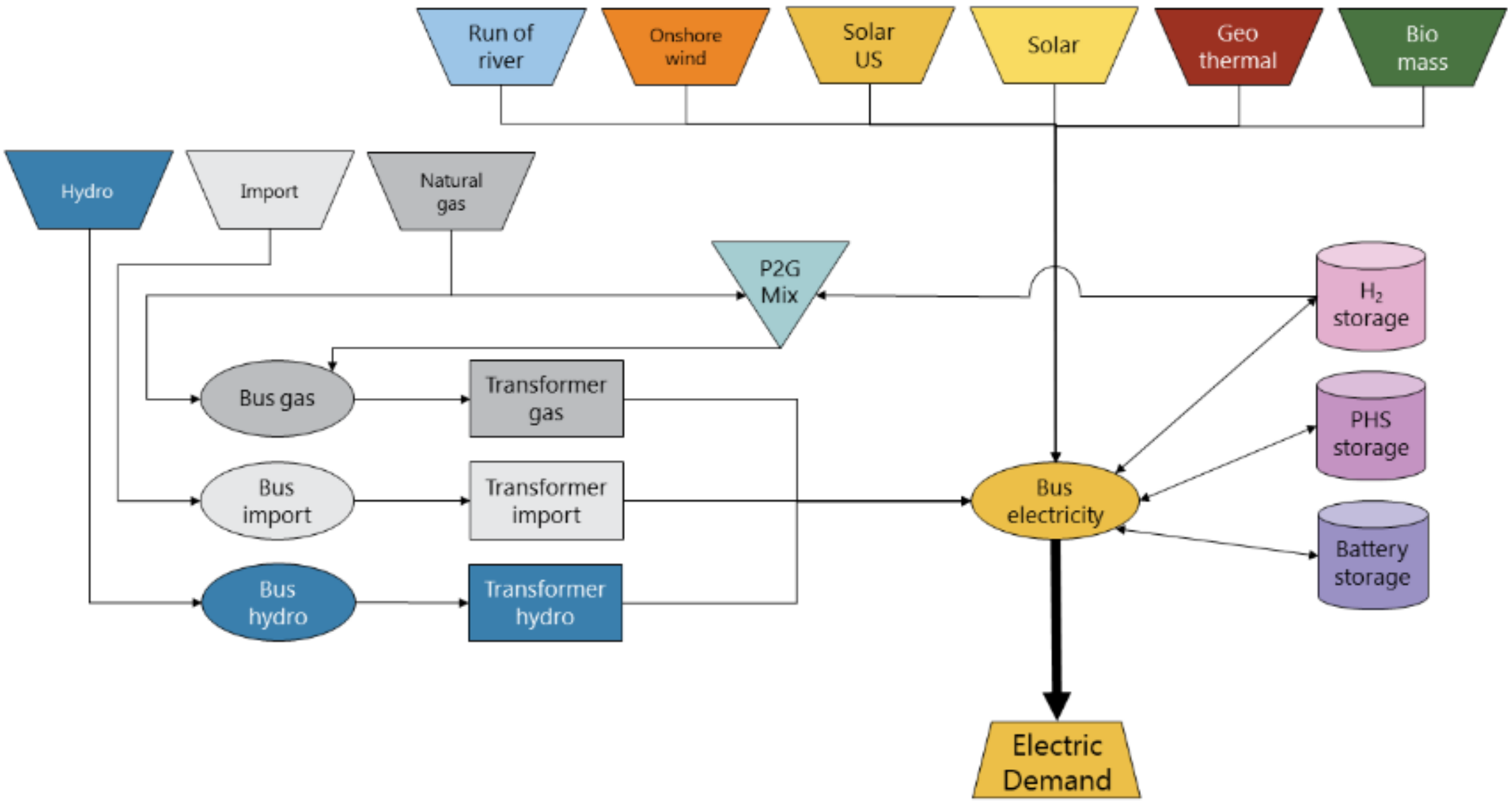

*Figure 5 - Oemof operational scheme for the Italian power sector in 2030*

For the expansion capacity optimization, in each market zone generation capacity of rooftop PV, utility scale PV and onshore wind can be increased, batteries, $H_2$ tanks, fuel cells and electrolyzers can be installed and powerlines linking nodes can be reinforced (details are outlined in Table 2). These technologis are the ones considered in the Italian decarbonization strategy for the power sector in 2030 [6]. The existing capacity for the power plants is taken from Terna [24], while for powerlines, it is obtained from the Development Plan [18]. Hydro power cannot be expanded in the model since its potential in Italy is limited by environmental reasons [39] and also bioenergy is expected to decrease in 2030 according to the Integrated National Energy and Climate Plan [6]; geothermal power is restricted to the Tuscany region and its enhancements are not considered due limited increases predicted in various Italian development strategies [40]. Offshore wind was not included in these scenarios, since so far its installation has encountered a large social resistance and very few MWs have received approval, mainly for experimental purposes [41]. Further work will also consider the expansion of this technology. Batteries and hydrogen storage technologies existing capacities are set to zero, since there are currently no relevant installations for centralized power generation in Italy, but they can be expanded by the optimizer. Maximum expansion capacities for rooftop PV, utility scale PV and onshore wind are provided by Trondle et al. [42], taking into account technological feasibility and public acceptance of renewable power plants. The investment costs for 2030 derive from various sources [26,43–46] and technical parameters for batteries [25,47,48] and hydrogen [49] are provided. The cost of expanding transmission lines is taken from Batas Bjelic and Rajakovic [50].

| *Market zone* | *Technology* | *Capacity [unit]* | *Maximum expansion [unit]* | *CAPEX [€/unit]* | *Unit ** |
|---|---|---|---|---|---|
| North | On-shore wind | 141.98 | 19522.2 | 1040 | MW |
| | Rooftop PV | 7991 | 207034.4 | 788.4 | MW |
| | Utility Scale PV | 1266 | 4449.6 | 350.3 | MW |
| | Battery | 0 | No limit | 158.9 | MWh |
| | Hydrogen tank | 0 | No limit | 10.0 | MWh |
| | Electrolyzer | 0 | No limit | 350.0 | MW |
| | Fuel cell | 0 | No limit | 339.0 | MW |
| | Transmission line to Centre-North | 4000 | No limit | 132550 | MW |
| | On-shore wind | 162.25 | 9450.4 | 1040 | MW |
| | Rooftop PV | 1660.1 | 36639.1 | 788.4 | MW |
| Centre-North | Utility Scale PV | 273.8 | 6755.9 | 350.3 | MW |
| | Battery | 0 | No limit | 158.9 | MWh |
| | Hydrogen tank | 0 | No limit | 10.0 | MWh |
| | Electrolyzer | 0 | No limit | 350.0 | MW |
| | Fuel cell | 0 | No limit | 339.0 | MW |
| | Transmission line to North | 1300 | No limit | 132550 | MW |
| | Transmission line to Centre-South | 2700 | No limit | 132550 | MW |
| Centre-South | On-shore wind | 2071.9 | 10025.0 | 1040 | MW |
| | Rooftop PV | 2316.9 | 76455.4 | 788.4 | MW |
| | Utility Scale PV | 1136.5 | 4090.2 | 350.3 | MW |
| | Battery | 0 | No limit | 158.9 | MWh |
| | Hydrogen tank | 0 | No limit | 10.0 | MWh |
| | Electrolyzer | 0 | No limit | 350.0 | MW |
| | Fuel cell | 0 | No limit | 339.0 | MW |
| | Transmission line to Centre-North | 1300 | No limit | 132550 | MW |
| | Transmission line to Sardinia | 1028 | No limit | 132550 | MW |
| | Transmission line to South | No limit** | No limit | 132550 | MW |
| South | On-shore wind | 4246.3 | 9620.0 | 1040 | MW |
| | Rooftop PV | 2557.2 | 29382.1 | 788.4 | MW |
| | Utility Scale PV | 822.9 | 15561.1 | 350.3 | MW |
| | Battery | 0 | No limit | 158.9 | MWh |
| | Hydrogen tank | 0 | No limit | 10.0 | MWh |
| | Electrolyzer | 0 | No limit | 350.0 | MW |
| | Fuel cell | 0 | No limit | 339.0 | MW |
| | Transmission line to Centre-South | 4600 | No limit | 132550 | MW |
| | Transmission line to Calabria | 1100 | No limit | 132550 | MW |
| Sardinia | On-shore wind | 1105.3 | 8138.9 | 1040 | MW |
| | Rooftop PV | 441 | 18666.5 | 788.4 | MW |
| | Utility Scale PV | 466 | 7240.8 | 350.3 | MW |
| | Battery | 0 | No limit | 158.9 | MWh |
| | Hydrogen tank | 0 | No limit | 10.0 | MWh |
| | Electrolyzer | 0 | No limit | 350.0 | MW |
| | Fuel cell | 0 | No limit | 339.0 | MW |
| | Transmission line to Centre-South | 1295 | No limit | 132550 | MW |
| Sicily | On-shore wind | 1904.1 | 8103.7 | 1040 | MW |
| | Rooftop PV | 949 | 31458.1 | 788.4 | MW |
| | Utility Scale PV | 482 | 4142.8 | 350.3 | MW |
| | Battery | 0 | No limit | 158.9 | MWh |
| | Hydrogen tank | 0 | No limit | 10.0 | MWh |
| | Electrolyzer | 0 | No limit | 350.0 | MW |
| | Fuel cell | 0 | No limit | 339.0 | MW |
| | Transmission line to Calabria | 1200 | No limit | 132550 | MW |
| Calabria | On-shore wind | 1125.8 | 3623.9 | 1040 | MW |
| | Rooftop PV | 396.8 | 11193.6 | 788.4 | MW |
| | Utility Scale PV | 141.0 | 644.5 | 350.3 | MW |
| | Battery | 0 | No limit | 158.9 | MWh |
| | Hydrogen tank | 0 | No limit | 10.0 | MWh |
| | Electrolyzer | 0 | No limit | 350.0 | MW |
| | Fuel cell | 0 | No limit | 339.0 | MW |
| | Transmission line to South | 2350 | No limit | 132550 | MW |
| | Transmission line to Sicily | 1100 | No limit | 132550 | MW |

*Table 2 - Decision variables set for each node of the model.*

**for each row, the unit of measurement is specified in the last column.*

***this powerline has no limit since, according to Terna, no bottleneck happened and it has an overly large capacity with respect to the actual flow*

## 4 Results and discussion

### *4.1 Full decarbonization and power self-sufficiency of the electricity sector in Italy*

The model of the Italian power sector explained in the previous section is employed to study what technologies would be needed to achieve 2030 targets and then reach a full decarbonization. The set of optimal solutions at different emissions reduction levels obtained is reported as a black line in Figure 6, obtained with an hourly resolution and employing all the decision variables previously described. A green star marker shows the baseline case (2019 scenario, considering coal power production with no emissions limitation) and a red star marker indicates the scenario that solely supposes coal phase-out (with data for 2030 but no possibility of RES expansion). These two cases are not included in the optimal solutions collection, which allows investments for the optimization variables and assumes coal phase-out. The emissions reduction goals for the power sector are depicted in the graph as vertical lines. They are the objectives posed by the European Green Deal [51] and the carbon budgets for the Italian electric system, set by the IPCC [1]. The latter is a total amount of emissions allowed to reach a certain goal (clarified in the legend of Figure 6). To calculate the admissible value of carbon dioxide emissions for 2030 we assume a linear decrease for yearly $CO_2$ emitted by the Italian power sector and that the sum until 2030 must be the same as the carbon budget, taking inspiration from [52].
A relevant contribution to reach decarbonization goals is offered by coal phase-out, followed by the installation of utility scale PV (detailed in Table 3), which is a very cost-efficient technology: combining coal power plants ruled out and capacity expansion allowed, it is possible to reduce the total $CO_2$ annual emissions by 30%. To stay below the 1.5°C threshold (purple line), a deeper transition is required, but in 2030 a 50% cost increase compared to the 2019 baseline (7.7 billion euros) would be enough. In the -93% emissions scenario total expenses would be doubled and for the full decarbonization they reach four times the reference value. It is essential to observe the costs of transition since they have an impact on society and families through energy prices. From Figure 6 it appears that the first steps of transition of the Italian power sector would need a limited rise in costs, but when getting closer to the total decarbonization expenses climb rapidly. Considering other options in the model could provide different points of view, for example integrating other energy sectors, including biomass or gas cycles with carbon capture, electric connection to other countries or expanding PHS, to cite some of the possibilities. Nuclear power has not been considered since two referendum banned it from the current Italian power system and according to the national development plans [6,53] it will not be installed in the future, even if this technology has a large potential in the energy transition. The overall message is that a full decarbonization of the Italian power sector in 2030 is feasible but at elevated costs within the framework outlined in this model, while achieving 2030 climate goals for the electricity sector can be done with limited investments. Policymakers should focus on achieving these intermediate objectives and prepare detailed strategies for a complete decarbonization of the power sector more shifted in the future, for example in 2040. In the next years, technology developments could make innovative solutions available and competitive, like carbon capture and storage (CCS) for gas power plants or biomass plants.

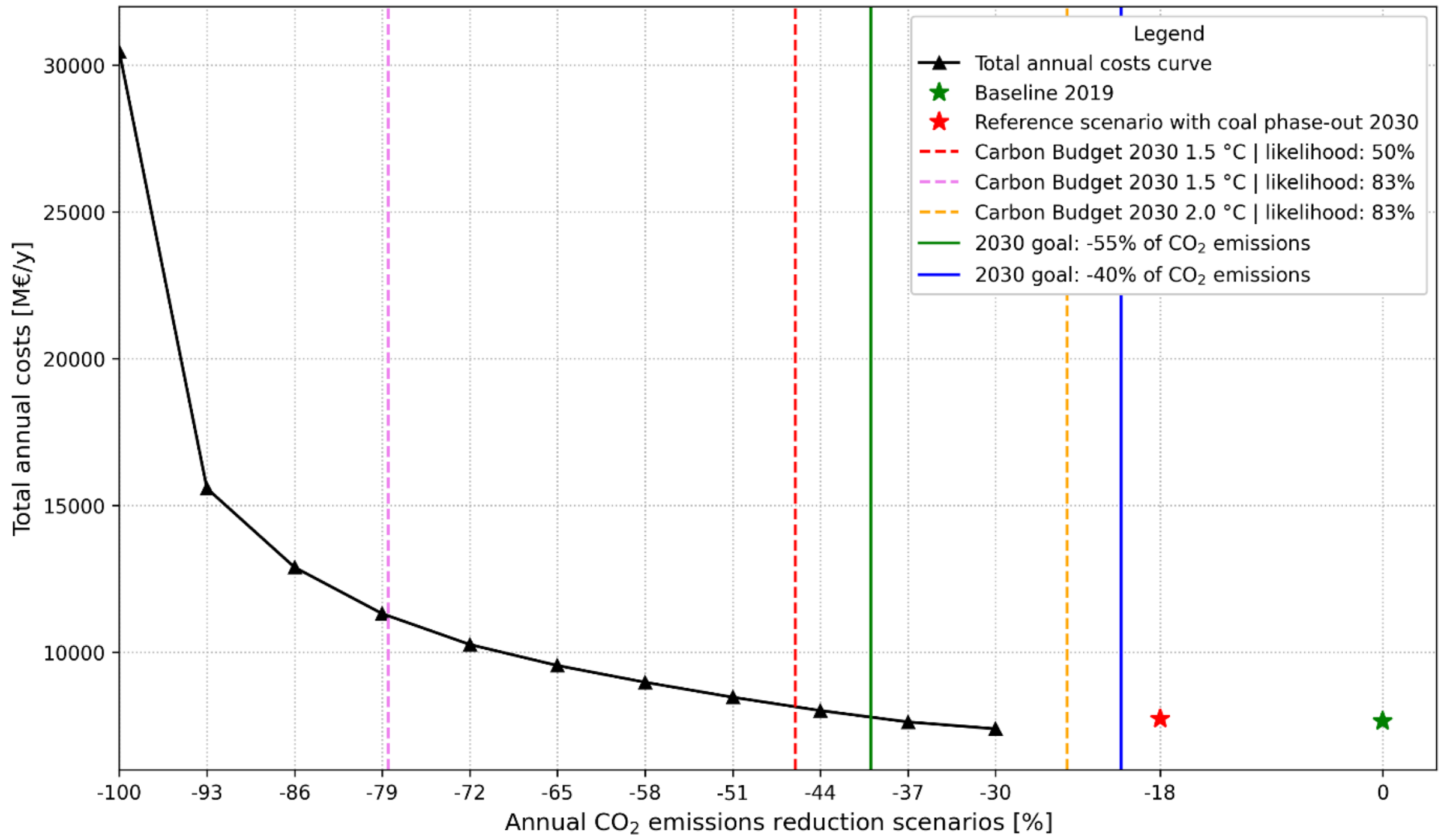

*Figure 6 – Optimal solutions for different emissions reduction levels and reference $CO_2$ emissions thresholds, All Variables Case*

Figure 7 better displays the allocation of costs at increasing degrees of decarbonization. Going to a more carbon-free system the expenditures for natural gas fuels are substituted by initial capital (CAPEX) to install new RES. The huge system cost increase at -100 % is mainly driven by rooftop PV, since utility-scale reaches full potential in the previous steps, li-ion batteries, hydrogen storage and powerlines expansion.

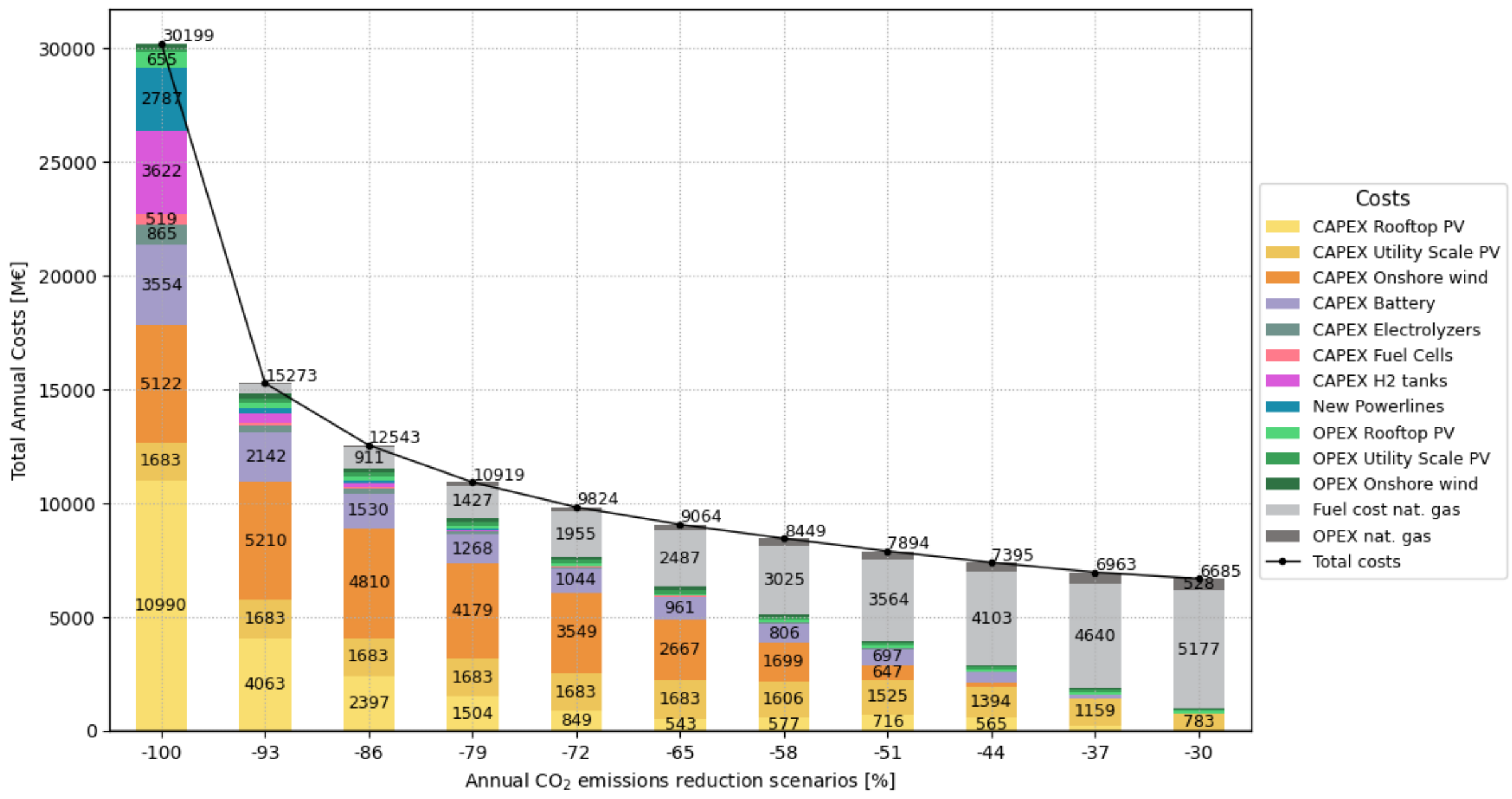

*Figure 7 – Total annual system cost curve at different emissions reduction cases, with cost items indicated in the columns for each case.*

Table 3 illustrates the output values of the installed power plants and storage technologies for the various scenarios at the national level, concordantly with the outputs in Figure 7. It is evident that solar utility scale is already convenient from an economic point of view and that synergies between wind and solar power become important for reaching great abatements. Only the last step for a completely carbon-free power sector requires an enormous amount of expanded capacity, explaining the great increase in costs shown in Figure 6 and 7, since removing completely fossil fuel driven power plants entails installing a great amount of RES technologies which are not exploited at their full potential. The optimization needs to install more capacity that can produce electricity in sunny hours and more storage to make it available in hours with a large request. These scenarios would need a deeper evaluation of the technical feasibility of the outlined solutions, for example considering the actual size of the required hydrogen storage or the availability of rare earths for the development of these technologies.

| *Scenario* | *On-shore wind [GW]* | *Solar rooftop [GW]* | *Solar Utility Scale [GW]* | *Li-ion batteries [GWh]* | *Electrolyzers [GW]* | *Hydrogen tanks [GWh]* | *Fuel Cells [GW]* |
|---|---|---|---|---|---|---|---|
| Existing | 10.8 | 16.3 | 4.6 | 0 | 0 | 0 | 0 |
| -30% | 10.8 | 16.3 | 24.5 | 0 | 0 | 0 | 0 |
| -37% | 10.8 | 19.6 | 34.1 | 13.8 | 0 | 0 | 0 |
| -44% | 12.2 | 23.8 | 40.1 | 32.2 | 0.2 | 2.2 | 0.1 |
| -51% | 17.0 | 25.8 | 43.4 | 48.7 | 0.9 | 8.1 | 0.2 |
| -58% | 27.0 | 23.9 | 45.5 | 56.4 | 1.1 | 10.1 | 0.3 |
| -65% | 36.3 | 23.5 | 47.5 | 67.3 | 1.1 | 10.9 | 0.3 |
| -72% | 44.8 | 27.5 | 47.5 | 73.0 | 1.5 | 25.1 | 0.4 |
| -79% | 50.8 | 36.2 | 47.5 | 88.7 | 2.5 | 68.5 | 0.7 |
| -86% | 56.8 | 48.0 | 47.5 | 107.0 | 4.2 | 253.8 | 1.5 |
| -93% | 60.7 | 70.0 | 47.5 | 149.9 | 6.1 | 567.2 | 4.0 |
| -100% | 59.8 | 161.5 | 47.5 | 248.6 | 18.2 | 4922.4 | 13.0 |

*Table 3 - Total installed capacity for different expanding technologies in various scenarios*

Figure 8 portrays the electricity mix in 2030 for different $CO_2$ emissions reduction scenarios for *All Variables Case,* so Case E. On the right, the Business-As-Usual 2030 case is represented (coal-fired power plants and no expansion of RES). Going towards a fully clean power sector, generation from RES increases, while natural gas consumption is reduced and import goes to zero since it accounts for fossil fuels used in other countries (in this case France) to produce electricity. Besides -93 and -100% emission cases, excess electricity remains a contained amount, no more than 10% extra generation (Table 4), thanks to flexibility provided by storage technologies and powerlines. Up to a 65% abatement (in line with the effort required to the power sector according to the European package Fit-for-55 [54]) excess energy represents less than 5% of the generated. A large increase in the ratio of generated energy over useful one is present for full decarbonization, due to the extreme added capacity for generation and storage. For the highest decarbonization levels, particularly large variable RES capacities produce more electricity than the one needed, but also the conspicuous utilization of transmission lines and storage implies major losses due to charging and discharging cycles. Solar generation increases from the first decarbonization steps since it is exceptionally economically feasible; when PV starts creating overgeneration, wind capacity is enlarged. For great abatement scenarios, rooftop PV production is boosted thanks to the large availability of spaces for this technology.

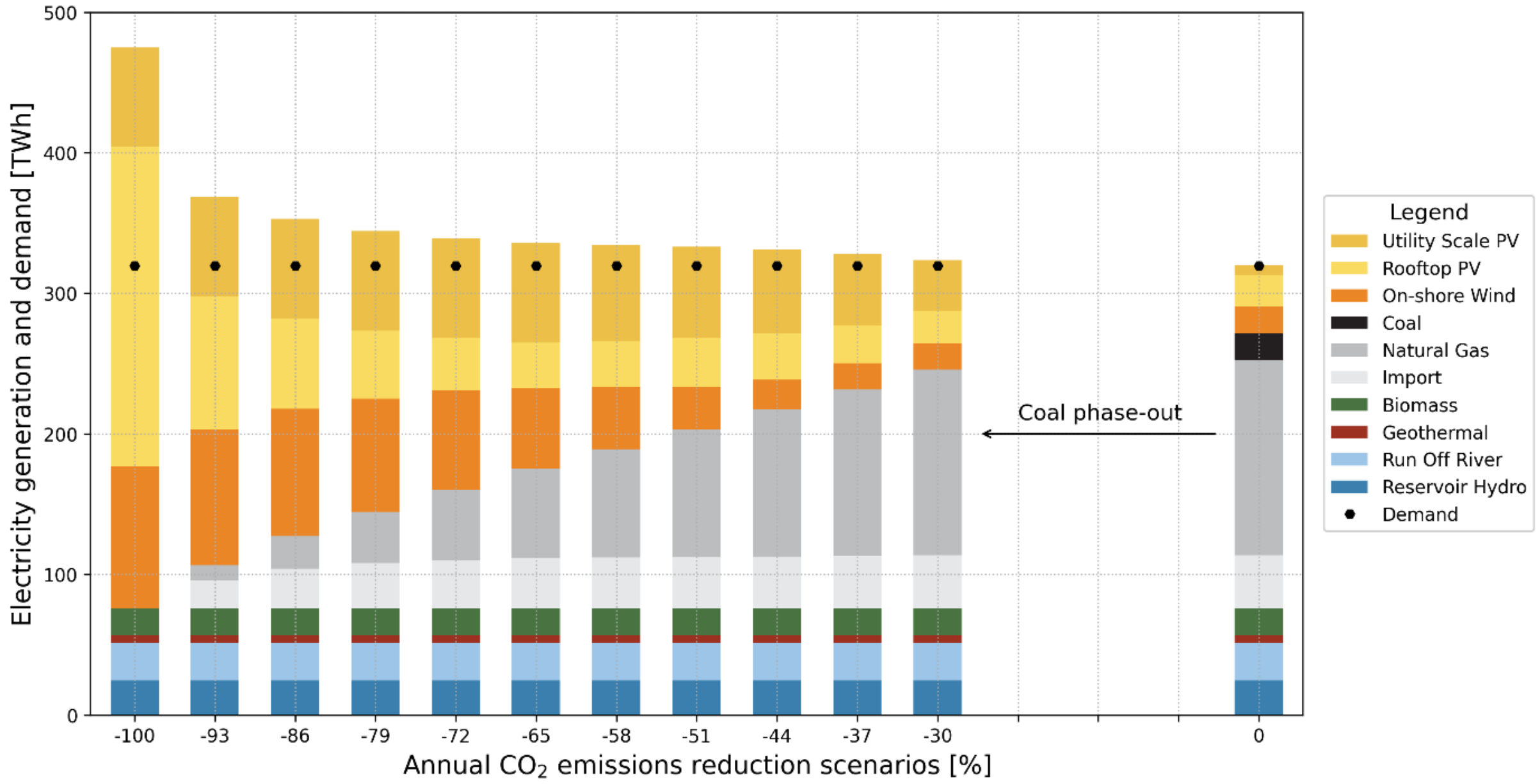

*Figure 8 - National electricity generation mix: different $CO_2$ emissions scenarios, All Variables Case*

| *Annual emissions reduction scenarios [%]* | *-100* | *-93* | *-86* | *-79* | *-72* | *-65* | *-58* | *-51* | *-44* | *-37* | *-30* |
|---|---|---|---|---|---|---|---|---|---|---|---|
| Generated electricity/useful electricity [%] | 149 | 115 | 110 | 108 | 106 | 105 | 104 | 104 | 104 | 103 | 101 |
| Losses for charge-discharge storage/useful electricity [%] | 8.8 | 6.6 | 4.8 | 3.6 | 2.9 | 2.4 | 2.1 | 1.9 | 1.2 | 0.7 | 0.3 |
| Losses for transmission/useful electricity [%] | 1.2 | 0.5 | 0.5 | 0.4 | 0.4 | 0.3 | 0.3 | 0.3 | 0.2 | 0.2 | 0.1 |

*Table 4 – Percentage of generated electricity and losses over useful one for different reduction scenarios*

Figure 9 shows the total capacity installed for RES that can be expanded by the optimizer (On-shore wind, Rooftop PV, Utility Scale PV) in each node for different reduction cases:

- Business-As-Usual in 2030, with coal still employed;
- the first scenario, assuming coal phase-out and expansion capacity but no constraint on the emissions, -30%;
- -44%, that as visible in Figure 6 allows to achieve the 2030 European goal of -55% emissions with respect to 1990;
- -79% case reaches the annual carbon budget to keep temperatures below 1.5°C;
- The step before full decarbonization in this study;
- total decarbonization for power sector (-100%).

Higher $CO_2$ reductions require a larger installation of variable RES technologies, with a remarkable expansion of rooftop PV in the North. In the fully clean power sector case, the capacity installed of this technology appears to be an extremely large value. Even if this analysis would require a deeper evaluation of technical and social feasibility, the latest trends in solar capacity installed globally seem to confirm similar numbers [55]. Finally, it is interesting to observe how wind power grows to complement solar power, besides in the final scenario where the former is dramatically substituted by the latter.

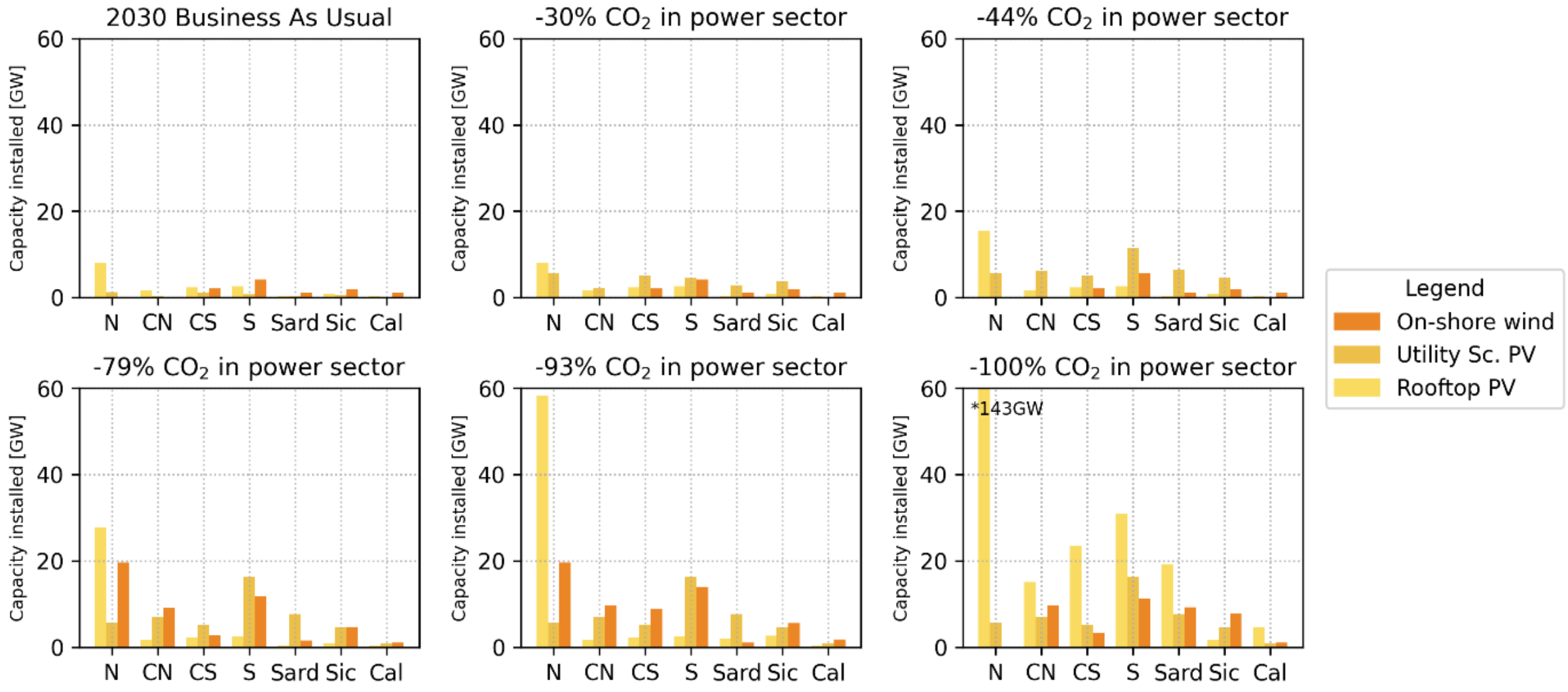

*Figure 9 – Total installed capacity of expanding generation technologies in each node for different scenarios, All Variable Case*

### *4.2 Relevance of batteries, hydrogen storage and transmission lines as optimization variables in energy modelling*

It is important to highlight the relevance of including technologies that can offer flexibility to the grid in power systems transition plans. The objective is to prove that it is hardly possible to set up transition plans for electricity sectors without allowing the expansion of some form of system flexibility. Figure 10 portrays the optimal solutions for the Italian power system at increasing $CO_2$ abatement with different sets of optimization variables:

- Case A which uses as decision variables: Rooftop PV, utility scale PV, on-shore wind
- Case B which uses as decision variables those of Case A plus transmission lines
- Case C which uses as decision variables those of Case B plus electrolyzer, FC, $H_2$ tank
- Case D which uses as decision variables those of Case B plus batteries
- Case E which uses as decision variables those of Case C plus batteries (*All Variables case*)

The graph underlines the crucial role of storage technologies and transmission lines in lowering the power sector cost for a defined $CO_2$ abatement level. In Case A and B, going beyond an 86% reduction is impossible without load shedding. These two cases are displayed only to highlight the role that balancing and storage solutions (in the other cases C, D and E) can have in decreasing the total system costs. In this analysis of the Italian electricity system decarbonization, it appears that batteries (mainly used for daily storage of energy) are more convenient than hydrogen storage (preferred for interseasonal storage thanks to its limited gas loss) [56–58]. Storage technologies can be the driving force towards a low-carbon electricity system. Whenever renewable electricity is higher than the demand in one time period, it is possible to store it and employ it when the system needs it, instead of simply curtail it. This approach helps reduce the required capacity of RES.

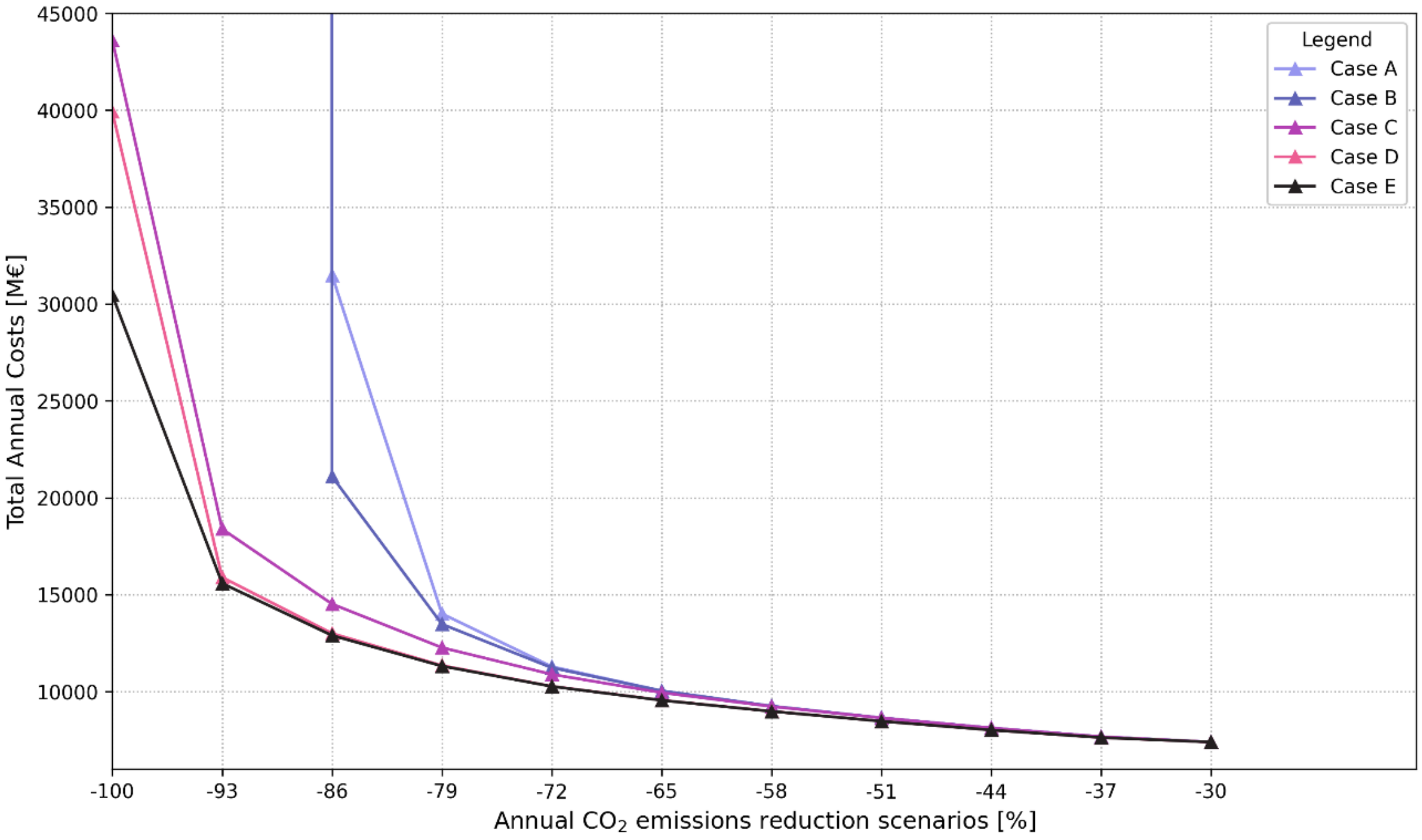

*Figure 10 – Optimal solutions at different emissions reduction levels for different decision variables sets*

** Costs for case A and B for -93 and -100% cases skyrocket to values in the order of $10^6$ million euros, they are thus excluded from the graph*

Figure 11 illustrates the optimal capacities installed for li-ion batteries and hydrogen energy storage each market zone for 2030 BAU, -44%, -65% -79%, -93% and -100% $CO_2$ emissions. PHS capacities are the same in every scenario since the expansion of this asset is not allowed in the model. In the -44% case, a limited capacity of li-ion batteries is visible in the South and Sardinia; at -79% hydrogen tanks are deployed in the South. The amount of hydrogen required for the complete decarbonization of the power sector is outstanding, particularly when compared to the prior -93%, showing a more reasonable capacity. This can create issues in terms of technical feasibility of the complete decarbonization in this framework. A more detailed evaluation of other options, for example CCS, sector coupling and demand shifts, could offer the opportunity to decarbonize at a much lower installed capacity and costs. Demand side management could be particularly effective in the North, since the enormous installation of hydrogen storage could be substituted by small shifts in the hourly consumed electricity during peak hours [59–61].

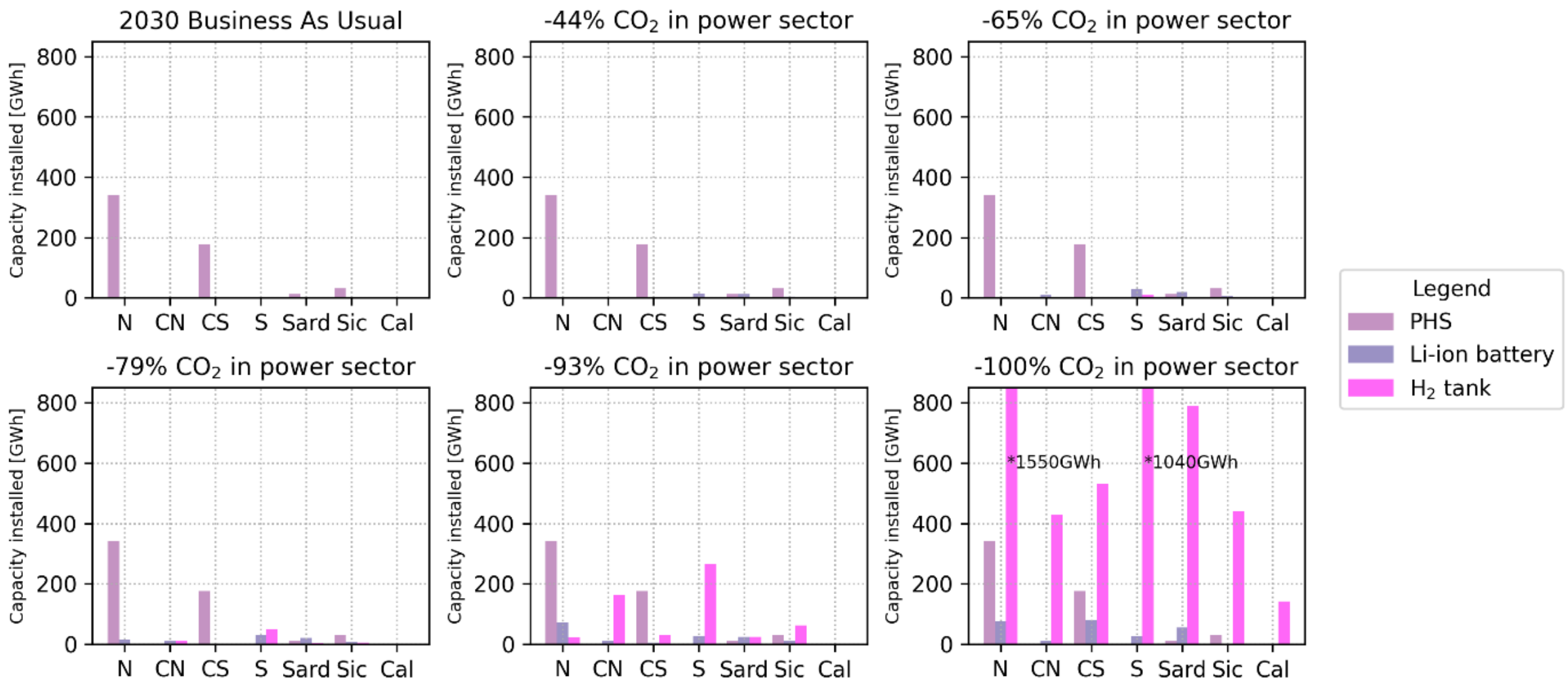

*Figure 11 – Total installed capacity of storage technologies in each node. 2030 baseline with coal power plants (left), coal phase-out with no $CO_2$ constraint case (centre), electricity sector carbon neutrality (right), All Variables Case*

Transmission lines allow energy to flow from one region to another, exploiting excess electricity in one and avoiding shortages in another. The optimization outcomes (Figure 12) illustrate a relevant expansion of the powerlines for the total decarbonization of the power sector. This extension goes especially from southern nodes, where there is a notable RES potential, to the Northern ones, which presents a remarkable demand (9 GW from Centre-North to North and 8 GW from Centre-South to South). In the complete decarbonization case, the cost for their expansion is around 9% of the total system annual cost. However, for an actual implementation, social acceptance should be considered and could lead to larger expenditure, for example to build underground lines.

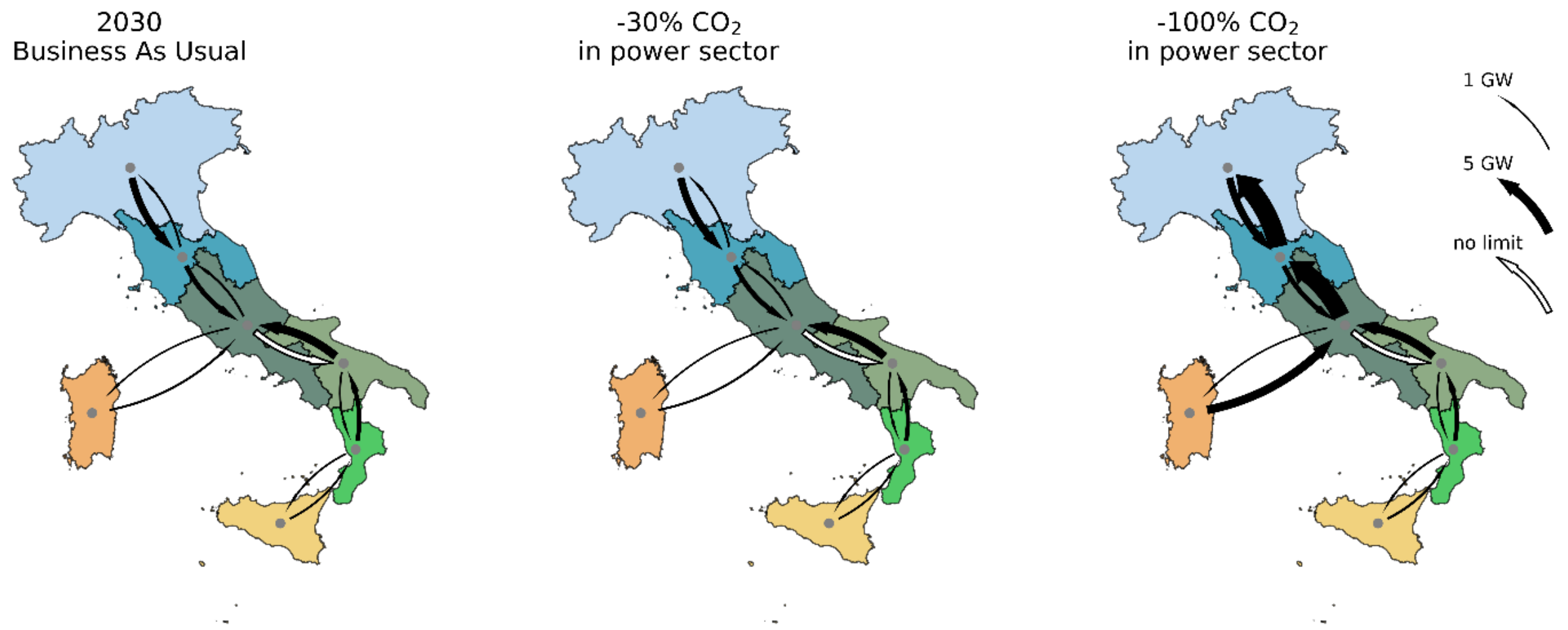

*Figure 12 - Transmission lines in 2030 baseline with coal power plants (left), coal phase-out with no $CO_2$ constraint case (centre), electricity sector carbon neutrality (right), All Variables Case*

*4.3 Sensitivity analysis on the share of hydrogen blended into gas pipelines*

The upper graph in Figure 13 depicts the outputs of the optimization with a share of hydrogen in gas pipelines of 20%, in agreement with [62] , the lower one of 50%, following instead NREL [63]. The latter percentage would need some enforcement of the pipelines. This can be an expensive operation, but it can be a beneficial trade-off to increase the utilization of the hydrogen infrastructure that in the future could be already installed for storage or to produce fuels. Other references advice diverse shares, so no clear accordance is highlighted about this parameter. The pink bars in the graphs represent the yearly energy content of hydrogen sent to fuel cells, while the grey parts stand for the annual $H_2$ energy inserted in the gas grid. The line shows the share of hydrogen employed for the P2G over the total amount of $H_2$ produced. In the bottom case, a decent improvement in the exploitation of P2G technology is observable, but this result does not seem outstanding, even if it opens the way for new opportunities of exploitation for this hydrogen. This energy vector diffusion is hindered by an elevated cost of investment, but it can be a decisive variable in the achievement of a zero-carbon emissions energy sector. Thus, it is cardinal to widen the useful ways to employ it, that can be various. Another application for the P2G option is for new hydrogen demand, as indicated by Colbertaldo et al. [58]:it can be a proficient solution to satisfy the future demand for Fuel Cell Electric Vehicles (FCEV), by employing $H_2$ production from electrolyzers. The P2G component implemented in the model can be applied to other further optimizations to evaluate its convenience and suitability for future usages and scenarios for decarbonizing other sectors, like hard-to-abate industry and transport.

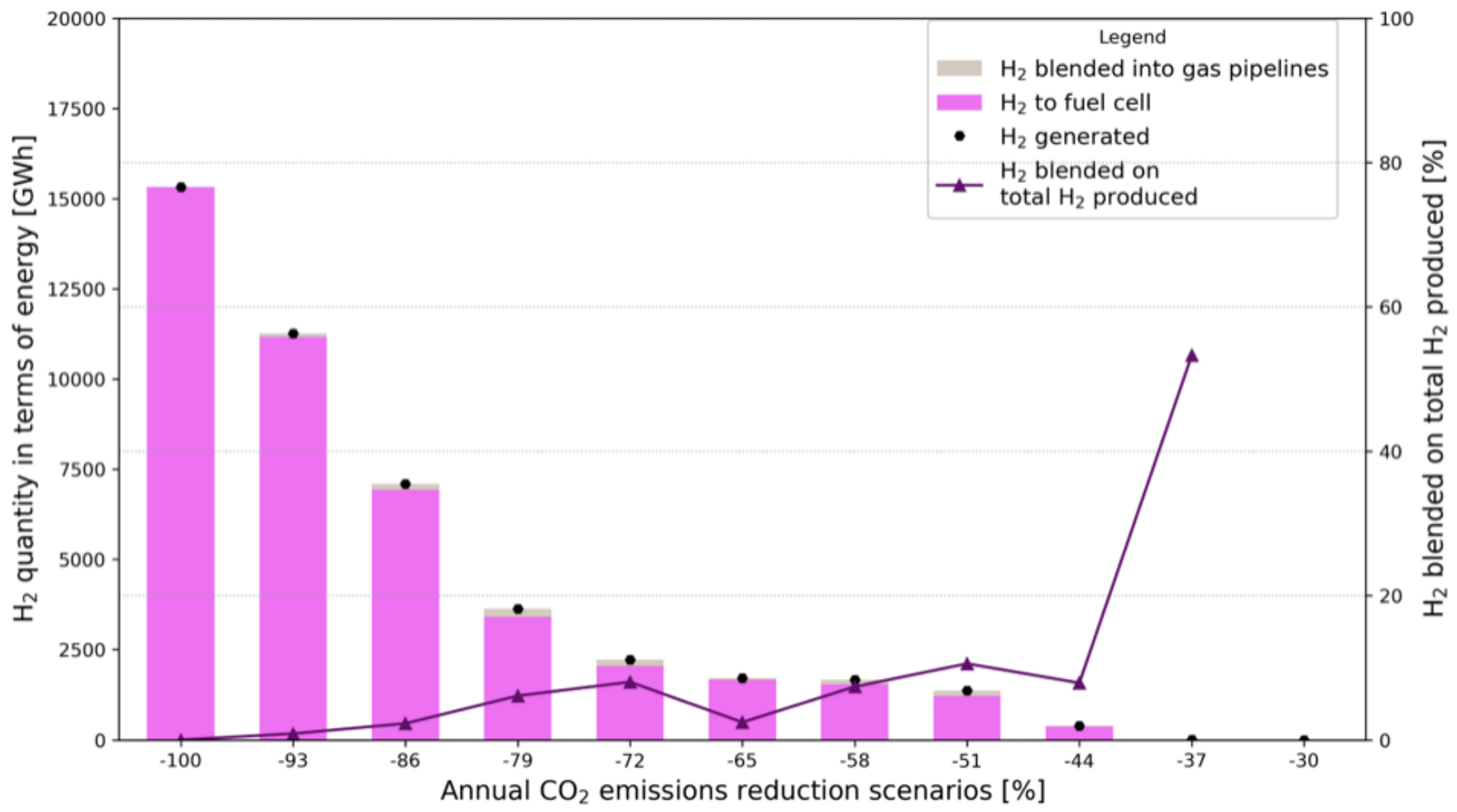

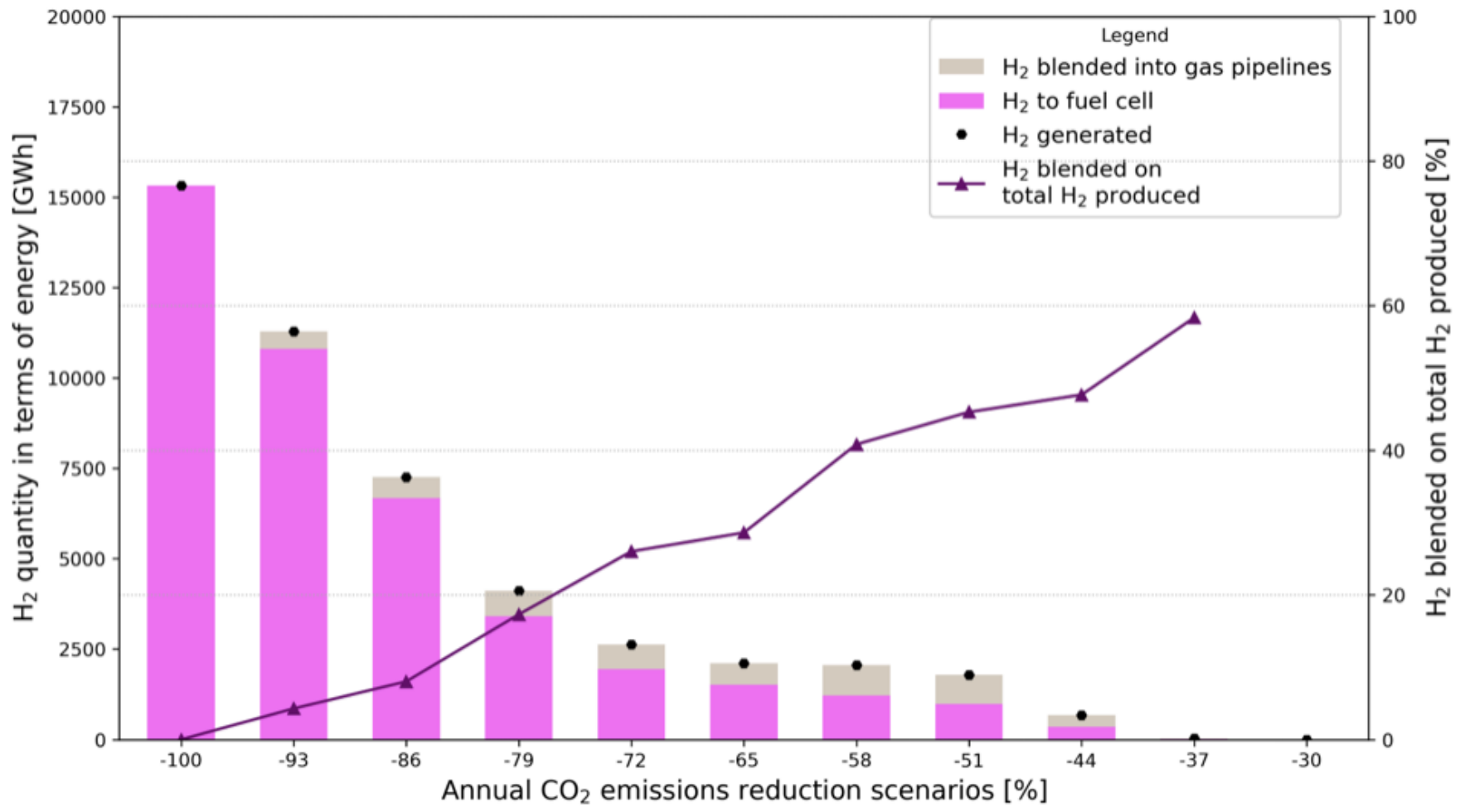

*Figure 13 - Power-to-Gas. $H_2$ share in natural gas pipelines, All Variables Case*

*Top figure: share of hydrogen in pipelines is 20%*

*Bottom figure: share of hydrogen in pipelines is 50%*

## 5 Conclusions and Policy Implications

In this work, the main objective was to investigate the modification of the Italian energy mix in 2030 when lowering $CO_2$ emissions till reaching carbon neutrality and achieving European decarbonization goals. The implemented approach offers a 7-nodes model with an hourly resolution which considers a high number of decision variables to perform expansion capacity optimizations.

The main message to decision makers is that coal phase-out and increased utility-scale solar emerge as low-hanging fruits for initial emissions reductions, achieving a 30% CO2 cut with almost no cost increase for the system. Also, transition away from coal-based power generation and embracing utility-scale PV technology are fully in line with Italy's national energy strategy. Coal phase-put alone would reduce $CO_2$ emissions by almost 20% with respect to today values (as in Figure 6), a relatively simple solution that is widely accepted by society and sustained by international institutions. Even without this fossil fuel in the Italian electricity mix, natural gas could still be kept in the first phases of the transition, offering flexibility and reserve margin. Then gas-fired power plants could be smoothly decommissioned until 2050, while providing the system with technologies that can replace their role as providers of programmable energy. Furthermore, utility-scale PV fields are economically more convenient than other renewable energy technologies in Italy thanks to the economy of scale of installing a large amount of MWs. Nowadays this technology is market-ready and can give a substantial push to emissions reduction without major costs enhancements for the system and thus the consumers. Other variable RES technologies must complement large scale PV installations: rooftop photovoltaics can be particularly useful to improve local distribution and generation of electricity, while wind power can help stabilize a system with a high share of electricity produced by solar technologies. The fiscal implications of this transition are deemed imperative, given their direct impact on societal dynamics and energy pricing. This study demonstrates that the initial stages of the transition experience relatively gradual cost escalation.

Further reductions become progressively more expensive, with the marginal cost of abatement rising exponentially as the emissions target approaches zero. In the progress towards total decarbonization, a more profound transformation is necessary: in this study, the final step before zero emissions would see the total system costs doubling, while a system fully decarbonized would cost almost four times compared to the current one. Massive buildouts of renewable capacity, storage and transmission are projected under deep decarbonization. Achieving net zero emissions requires boosting variable renewable generation to 200 GW above current levels. The technical feasibility of implementing such a rapid and large-scale transition by 2030 is highly uncertain. Challenges include materials availability, infrastructure upgrades, social acceptance, and electricity system stability with very high variable generation.

This suggests that policy makers would have to evaluate thoroughly all the possible solutions, cooperating with neighbouring countries to fight this global challenge. It is important to consider also the opportunities offered by the inclusion in the electricity mix of programmable technologies with net zero emissions, like nuclear or biomass plants with CCS, that can offer stability and reduce the overall system expenses. The enormous RES and battery capacities resulting from this analysis in the total decarbonization case are also underutilized, since they might be installed to provide electricity or store it just for a few hours during the year. Electricity interconnection between States and programmable sustainable technologies can instead decrease the installed RES capacity needed. Another pivotal element that is often overlooked is the demand side: with the right fiscal incentive consumers can easily shift part of their load to hours when renewable sources are more available. This can offer a cheap and fast solution to elevate peak hours prices for the system operators. Considering the results in Table 4, it is clear that a notable amount of excess electricity will be produced, leaving open many opportunities for synergies with other energy sectors that could exploit cheap electricity to produce green hydrogen or e-fuels. An integrated multi-sector perspective can unlock cheaper mitigation pathways than the modeled power sector focus. Heating, transport and industry decarbonization must also be addressed, leaving room for future improvements of the current study. An exhaustive analysis from decision makers should evaluate the whole energy system, to better balance the pros and cons of the transition in each one of them. Finally, since an increase in electricity prices might be inevitable, besides all the preventive measure taken, government should put in place social fiscal schemes to protect consumers. These should be progressive measure, offering larger incentive to the most vulnerable parts of society and ensure a decent quality and quantity of electricity to families and firms.

A reflection should be done also on the feasibility of the installed capacity resulting from the study: the enormous amount of new solar panels and batteries needed appear like a huge challenge for the future. This should not be underestimated, since it would require a large amount of materials, rare earths and electrical connections to be put in place. It is still to be reminded that this analysis has been done for 2030, while the full decarbonization is expected to happen 20 years later, giving more time to install these needed capacities and build up the system. While the economic and technical viability of fully decarbonizing the Italian power sector by 2030 is doubtful, more gradual multi-decade pathways aligned with technology learning curves may balance environmental and economic sustainability. Driven by incentives and government push, solar PV historical installations have grown way more that the projected values [55,64,65], thus a carefully thought policy framework can be a valuable factor to achieve a great transformation of the power sector. Near-term policies should prioritize low-cost, socially acceptable steps like phasing out coal and deploying utility-scale solar. Premature targets for full decarbonization could jeopardize public support.

This work shows the fundamental role of storage and flexibility options: batteries, hydrogen storage and powerlines are crucial to reduce overgeneration, particularly for high abatement scenarios, balancing the daily load with batteries and the seasonal differences with hydrogen. Reaching complete decarbonization would be economically impossible without increasing the storage and flexibility capacity available in the power sector. To achieve 2030 European goals, their development would be still moderate, but it is fundamental that policymakers include storages like li-ion batteries and hydrogen in their transition strategies, even for the first stages. This could help to test their integration in the system and experiment new types of electricity markets that can make storage margin capacity profitable. Investment in research and development are also paramount, in order to better study new technologies that seem promising for a carbon-free energy system in 2050.
Short-term energy storage solutions, such as batteries, can play a fundamental role in the contemporary power sector by addressing the intermittency challenges inherent to variable renewable energy sources like wind and solar. The integration of RES into the grid has introduced fluctuations in power generation due to the dependence on weather conditions. Batteries, with their rapid response times and ability to store and release electricity swiftly, provide a vital mechanism for maintaining grid stability. They ensure a consistent supply of electricity, minimizing the fossil-fueled backup power generation. Conversely, long-term energy storage, as exemplified by hydrogen in the paper, plays a distinct but equally significant role in the energy transition. Hydrogen tanks can accumulate copious amounts of energy in gas form during resourceful months when renewable sources are abundant, such as sunny summer days or windy periods. This surplus electricity is into hydrogen through electrolysis and stored for extended durations, effectively acting as an energy reservoir. Hydrogen's capacity for long-term storage addresses the seasonality of renewable resources, complementing short-term storage solutions. The data presented in Figure 10 indicates that in the context of a short-term transition, the incorporation of these storage solutions into the analysis yields relatively constrained benefits. However, when considering profound emissions reduction objectives, there is a discernible cost advantage associated with their implementation.

Another aspect that must be kept in mind for Italy is that the country exhibits significant disparities in both electricity generation profiles and power load demand, due to its considerable longitudinal extent. This geographic feature renders Italy unique, with distinct characteristics between its northern and southern regions that can pose an ulterior challenge: most of its national electricity demand is located in the North, while renewables sources are more abundant in the South, with longer sun and wind equivalent hours. Considering Italy as a single node in a model and thus offering a uniform climate policy for the whole nation could lead to non-optimal solutions.
Expansion capacity models optimize based on costs, thus considering the efficiency of transmitting electricity from the South to the North it might be more convenient to directly install technologies in the North, as evident in this case in Figure 9. This might collide with reality, since there is no single, perfect-foresighted owner of all the power plants that wants to optimize their portfolio, but many firms looking at their own profit. Many renewable technologies companies, in fact, are planning to locate new installation in the South, where capacity factors are higher and they can sell more energy [41]. In this context, decision makers must find a compromise between the cost-optimal solution for the country and the energy companies need to make a profit, both by creating a financial incentive to install RES capacity where necessary and by reinforcing the high voltage grid to favor electrical exchanges within the peninsula.
The study, in its current scope, does not delve into the intricacies of the distribution grid, and therefore, it does not encompass the nuanced issues associated with this aspect. The distribution grid represents a critical interface between localized power generation and end-users, with profound implications for grid management, resilience, and the overall integration of renewable energy resources. Hence, it is imperative to augment the distribution capacity and extend it geographically, to mitigate the escalating incidence of power outages and blackouts, that can hinder societal wellbeing. These disruptions are partially attributed to the mounting frequency and severity of extreme weather events due to climate change [66]. The effects of these global challenge are also impacting electricity demand and power generation, thus a deeper understanding of this phenomena is pivotal to build a reliable and resilient future power sector [67,68].

The adoption of modelling techniques within the context of energy systems presents a commendable approach for policymakers, offering structured means to consistently scrutinize their decarbonization initiatives. Through the utilization of optimization or simulation models, decision makers can thoroughly evaluate the reliability and effectiveness of various aspects of the energy system. This approach facilitates the validation of strategies instituted for energy transition, while bolstering the system's resilience against unforeseen perturbations. In the quest for successful decarbonization plans, the meticulous selection of temporal resolution emerges as a pivotal consideration. This holds particularly true in the case of power systems characterized by a substantial penetration of renewable energy sources. The adoption of a temporal resolution of at least one-hour timestep becomes imperative as it offers a more comprehensive understanding of the intricate interplay between meteorological conditions, electricity production and energy demand fluctuations. By examining these variabilities with such granularity, politicians are better equipped to formulate strategies that are adaptable to the dynamic nature of the energy landscape. The power system serves as enabler for an effective profound transition across the entire energy economy. As electric vehicles, heat pumps, and electrified heating processes continue to gain prominence, future electricity demand will increase and the hourly shape will change based on different necessities of transport, buildings and industry. A comprehensive and exhaustive analysis of power decarbonization pathways is thus an indispensable factor of any energy decarbonization strategy.

In conclusion, the study has also offered valuable insights into the implementation of Power-to-Gas technology, raising pertinent questions and opportunities for future research endeavors. Particularly noteworthy is the question surrounding the integration of hydrogen into natural gas pipelines, a topic that continues to captivate the scientific community's attention. This unresolved issue underscores the complexity and multifaceted nature of such an approach, necessitating further investigation and exploration. Consequently, the study's findings underscore the salience of delving deeper into the intricacies of $H_2$ incorporation into existing gas infrastructure, presenting a promising avenue for future research endeavors.
A limitation of this study is that it considers only the power sector, while the integration of heating and transport sectors could reduce the need for storage that drives up costs in the most extreme decarbonization scenarios. In fact, there is evidence that exploiting synergies between energy sectors and other forms of flexibility in other sectors (like thermal storage, vehicles batteries with smart charging and V2G) reduce the need for electrical storage [15,69,70]. Another limit is that this study does not explore in detail the practical feasibility of large capacities installations for drastic emission reduction scenarios, but we only ensure that they stay below a social potential threshold.

**CRediT authorship contribution statement**

A. Di Bella and F. Canti: Data curation, Formal analysis, Investigation, Methodology, Software, Visualization, Writing – original draft, Writing – review & editing.
M. G. Prina: Conceptualization, Writing - Review & Editing, Visualization, Supervision.
V. Casalicchio: Formal analysis.
G. Manzolini: Conceptualization, Project administration.
W. Sparber: Project administration, Funding acquisition.

**Declaration of competing interest**

The authors declare that they have no known competing financial interests or personal relationships that could have appeared to influence the work reported in this paper.

**Acknowledgements**

The authors thank Politecnico di Milano and Eurac Research for the opportunity to develop this study.

**Data availability**

Model is available at https://github.com/cerealice/oemof_Italy.

**Appendix A. – Electric demand calculations for the 7 zones configuration**

In this study, while minimizing the total system costs $C^{total}$, a constraint ($\epsilon_i$) for carbon dioxide emissions $CO_2^{total}$ [Mton$CO_2$/y] is imposed and iteratively reduced, obtaining each time an optimal power system at a different emissions level (Eq. A.1). The limit imposed goes from the absolute value of emissions of the model expansion capacity optimization without constraints to zero, decreased with a -10% step, thus exploring issues and opportunities of the transition to a fully decarbonized Italian power sector. The investment optimization with no limit already reaches a 30% emissions reduction with respect to the $CO_2$ yearly emitted today, thus the $\epsilon_i$ values are the ones outlined in Eq. A.2.

$$Min\ C^{total} \quad while\ CO_2^{total} \leq \epsilon_i \tag{A.1}$$

$$\epsilon_i \in current\ emissions \cdot [-30\%, -37\%, -44\%, -51\%, -58\%, -65\%, -72\%, -79\%, -86\%, -93\%, -100\%] \tag{A.2}$$

The total value of $CO_2$ emissions is calculated as reported in Eq. A.3.

$$CO_2^{total} = \sum_{t \in T} \sum_{n \in N} \sum_{s \in S} P_{t,n,s}^{fossil} \cdot co_{2,s}^{factor} \leq \epsilon_i \tag{A.3}$$

t = analyzed time step, from 1 to T
n = node considered, from 1 to N
s = generation source, from 1 to S

$P_{t,n,s}^{fossil}$ = thermal power supplied by the generator s in node n at time step t

$co_2{}^{factor}_{s}$ = emission factor for technology s in MtonCO2/MWh
$\epsilon$ = limit for the total amount of $CO_2$ emitted in total by the system. This is reduced at each iteration until being null

The total system annual costs $C^{total}$, that must be minimized, are calculated as in Eq. A.4.

$$C^{total} = C^{op} + \sum_{n \in N} (\sum_{s \in S} C_s \cdot P_{n,s}^{add} + \sum_{st \in ST} C_{st} \cdot E_{n,st}^{add} + \sum_{p \in P} C_p \cdot P_p^{add}) \tag{A.4}$$

st = storage technology, from 1 to ST
p = existing powerline connecting two zones, from 1 to P

$C_s$= resource s expansion capital cost
$C_{st}$=storage tech. st expansion capital cost

$C_p$=powerlines expansion capital cost
$P_{n,s}^{add}$= capacity added for source s in node n
$E_{n,st}^{add}$ = capacity added for storage st in node n
$P_p^{add}$ = capacity added for powerline p

The cost of operation ($C^{op}$) of the power system are expressed in Eq. A.5.

$$C^{op} = \sum_{t \in T} \sum_{n \in N} \sum_{s \in S} E_{t,n,s} \cdot vc_{n,s} \tag{A.5}$$

$E_{t,n,s}$= electricity generated by source s in node n at time step t

$vc_{n,s}$= variable cost of generation for the energy source s in node n

Each node n in each time step t has an electricity demand $D_{n,t}$ that has to be satisfied with the total electricity generated $E_t$ electricity at the time step t, also considering the charge $E_{t,n,st}^{charge}$ and discharge energy $E_{t,n,st}^{discharge}$ of the storage technology st in node n and the transmission losses $E_{t,p}^{trans\ loss}$ for every powerline p. It is possible to have an excess of generation $E_{t,n}^{excess}$ in each node n. It is mathematically described in Eq. A.6 and must be valid for each t ∈ T.

$$\sum_{n\in N}\sum_{s\in S}\sum_{p\in P}\sum_{st\in ST}\left(E_{t,n,s}+E_{t,n,st}^{discharge}-E_{t,n,st}^{charge}-E_{t,p}^{trans\,loss}\right)=\sum_{n\in N}D_{t,n}+E_{t,n}^{excess} \quad (A.6)$$

The power supplied by each generator unit must be positive and lower than its nominal power, as in Eq. A.7. Power provided by RES, is instead normalized by the resource availability, as in Eq. A.8.

$$0 \leq P_{t,n,s}^{fossil} \leq NP_{n,s} \quad (A.7)$$

$$0 \leq P_{t,n,s}^{RES} \leq NP_{n,s}\cdot a_{t,n,s}^{RES} \quad (A.8)$$

$NP_{n,s}$= nominal power of source s in node n

$a_{t,n,s}^{RES}$= availability of RES source s in node n at time step t

Finally, storage content and power flow through transmission lines are limited by their nominal values (Eq. A.9 and A.10 respectively).

$$SC_{t,n,st} \leq NP_{n,st} \quad (A.9)$$

$$PF_{t,p} \leq NP_{p} \quad (A.10)$$

$SC_{t,n,st}$= storage content at time step t in node n for storage technology st
$NP_{n,st}$= nominal storage capacity in node n for storage technology st

$PF_{t,p}$= power flowing in powerlines p in time step t
$NP_{p}$= nominal exchange capacity of powerlines p

**Appendix B. – Electric demand calculations for the 7 zones configuration**

For each region, a peak value and a time series of the hourly request are considered. Time series are obtained by dividing the hourly demand by Terna [31] and the maximum demand for the specific node. Some data is missing for the 7 zones case since it used to be aggregated differently. Regarding the Calabria region, the time series for the demand remains the same as in the South, but a new peak value must be calculated. The total demand for each region is known [19], which has to be equal to the peak value multiplied by the time series value for each time step. Applying Eq. B.1 to the Goal Seek function in Excel, peak load values for South and Calabria nodes are found.

$$D_{n}^{annual}=Peak_{n}\cdot\sum_{t=1}^{T}d_{t,n} \quad (B.1)$$

$D^{annual}$ = annual demand for each node n
$Peak_{n}$ = peak load for each node n

$d_{t,n}$ = demand time series value (between 0 and 1) for time step t

The same process is implemented for Centre-North and Centre-South peak values since the Umbria region shifts from the former to the latter. In this case, the time series for the RES profiles of the Centre-South zone needs to be readjusted to consider that electricity generation from this node also comes from the installed power in Umbria. The new RES availability in the Centre-South zone is calculated as follows (cf. Eq. B.2). First, the hourly electricity generation from Umbria and Centre-South before the modification in 2021 (i.e. Lazio, Marche and Campania) is computed. The Centre-South region composition in 2019 will be called *R3 2019*.

$$E_{t,s} = P_s^{installed} \cdot a_{t,n,s} \tag{B.2}$$

$E_{t,s}$= electricity produced for each RES source s in each timestep t for Umbria and *R3 2019*

$P_s^{installed}$ = installed power (existing and added) for each RES source s for Umbria and *R3 2019*

$a_{t,n,s}$ = RES profile time series value (between 0 and 1) for time step t and RES source s for Umbria and *R3 2019*

After this first step, the time series for the new Centre-South node is calculated as follows in Eq. B.3:

$$new\ a_{t,s}^{R3} = \frac{E_{t,s}^{Umbria} + E_{t,s}^{R3\ 2019}}{P_s^{installed\ Umbria} + P_s^{installed\ R3\ 2019}} \tag{B.3}$$

$new\ a_{t,s}^{R3}$ = new RES profile time series value (between 0 and 1) for time step *t* and RES source *s* in Centre-South.

In Table B.1 the numbers retrieved or computed for the peak load demand in the two market zones configurations are displayed.

*Table B.1 - Peak demand for each zone in the two configurations*

| Node | Peak demand 6 zones [GW] | Peak demand 7 zones [GW] |
|---|---|---|
| North | 35.82 | 35.82 |
| Centre North | 6.48 | 4.24 |
| Centre South | 9.02 | 11.25 |
| South | 5.31 | 4.37 |
| Sardinia | 1.56 | 1.56 |
| Sicily | 3.51 | 3.51 |
| Calabria | - | 0.94 |

**Appendix C. – Temporal resolution sensitivity analysis**

Temporal resolution is a crucial aspect for energy modeling towards systems with a consistent penetration of variable RES and analyses on this parameter are not particularly diffuse in literature [71]. In this Appendix, for the evaluation of the temporal resolution, two approaches were deployed. The first one averages values of consecutive time steps (called *downsampling* by Pfeninnger [72]) and the second one uses a clustering method (in this work with a *k-means algorithm* taken from Casalicchio et al. [73]). Through the latter, 4 representative weeks are selected. The idea behind reducing the time resolution of models, particularly if they have a large number of nodes and variables is to diminish computational effort. This was not a critical point for electric models mainly based on fossil fuels sources, that do not rely on weather availability. Modellers used many approaches to simplify temporal granularity in energy system models, but these may not be adequate to replicate fundamental dynamics for power sectors with high RES penetration, as shown in the following results [74]. Through the averaging method, very imprecise outcomes are produced, as in Figure D.1: RES generation is highly affected by averaged values since they do not consider the hourly weather variation and, as specified in literature [75–77], leads to an excessive overestimation of RES needed capacity. The 4 weeks clustering yielded more accurate results for the first $CO_2$ abatement scenarios, while for more severe ones, it presents a substantial overestimation of the system costs due to its inability to model the chronological order of storage contents properly. From outputs of the spatial and temporal resolution comparison and from sources in literature, it appears that power system models can be rather sensitive, particularly to time granularity.

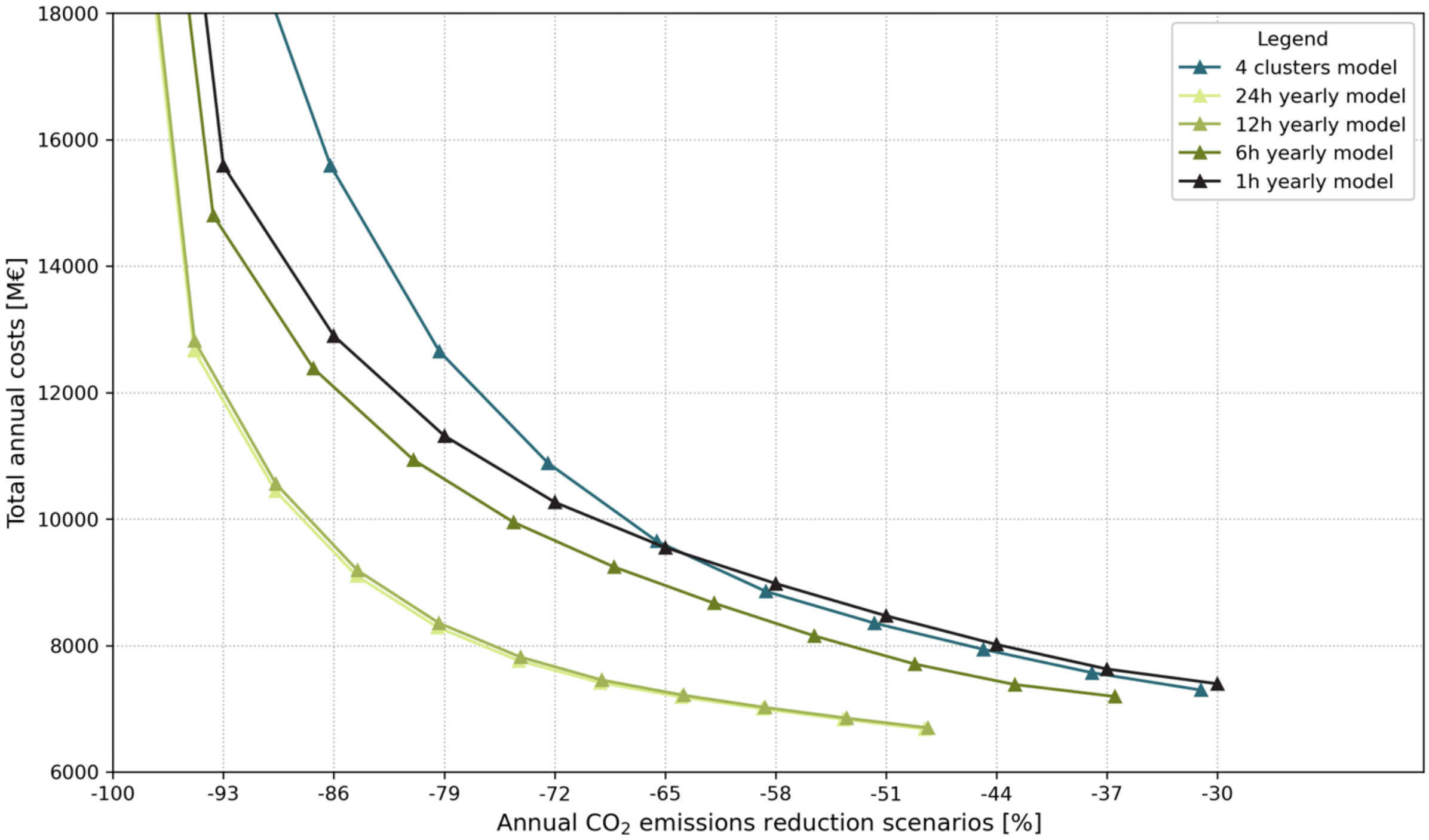

*Figure D.1 - Optimal solutions at different emissions reduction levels: temporal sensitivity analysis comparison of the two approaches*

The inability to represent storage correctly is evident in Figure D.2, where four tables representing the hourly trends for the same week are shown: on the left, electricity generation in the *All Variables case* (upper graphs) and for the optimization with 4 cluster weeks (down), on the right the storage content for the same cases respectively. The horizontal axis shows different numbers for the optimal and 4 clusters case since the first week (hours from 0 to 168) of the latter corresponds to a typical winter week in the former (hours 8402 to 8570). Trends are very different, particularly looking at the storage content evolution (here, negative values for discharging, positive ones for charging). Considering four separate weeks of the year and assembling them to simulate a whole year produces a loss of sequential order in the modeling of storage options [76]. According to Frew et al. [78], this issue must be handled with assumptions made by the modeler. An example of an approach to smooth this discontinuity, proposed in the paper, is to impose that the same storage content at the beginning and at the end of each day, or that the sum of the net storage flows goes to zero for each day.

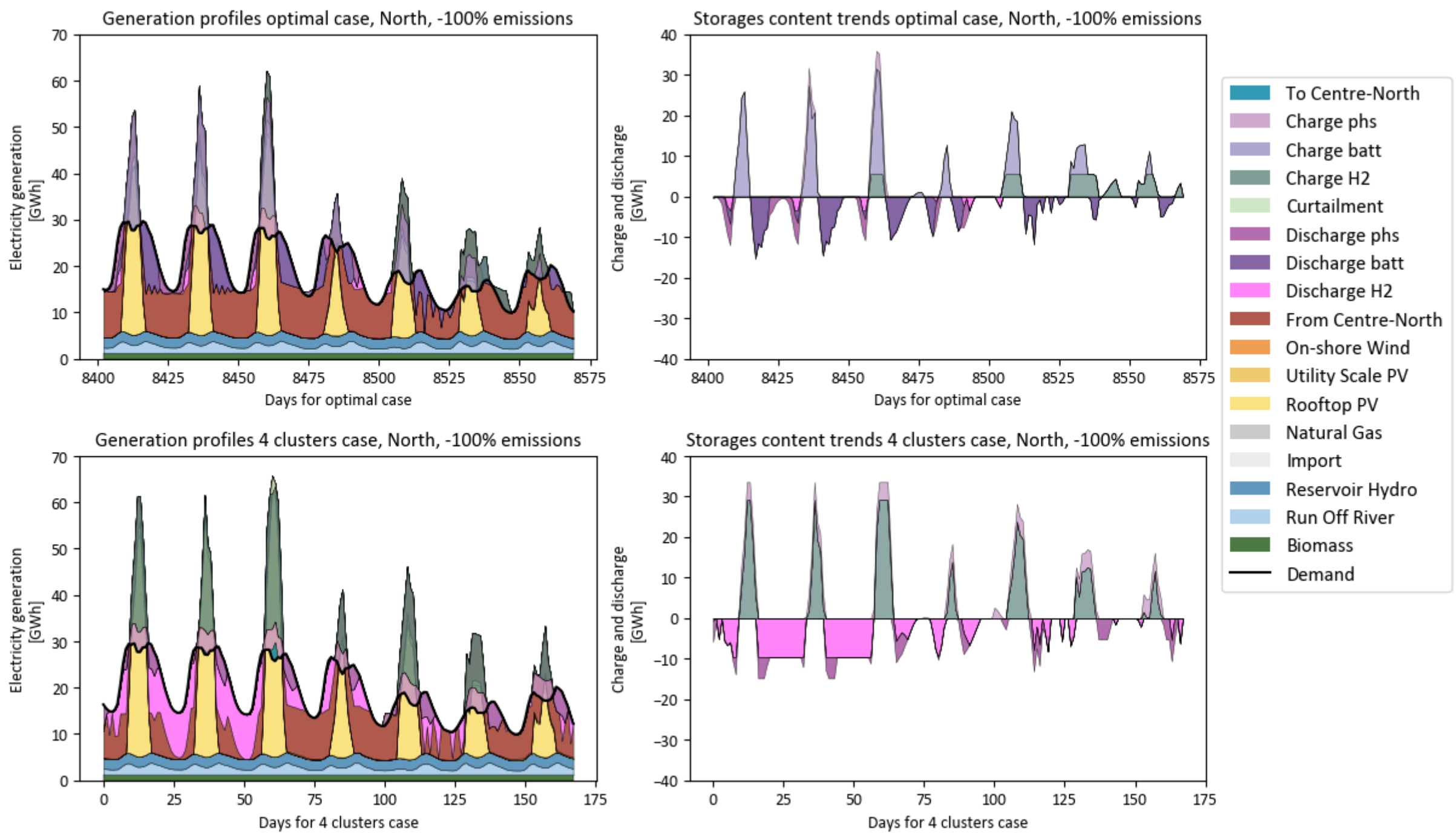

*Figure D.2 - Comparison between the Italian model and the clustered one, -100% $CO_2$ emissions scenario*

**Acronyms**

| | |
|---|---|
| BP | British Petroleum |
| capex | Capital Expenditure |
| CAES | Compressed Air Energy Storage |
| CCS | Carbon Capture and Storage |
| CHP | Combined Heat and Power |
| CO2 | Carbon Dioxide |
| CSP | Concentrated Solar Power |
| FC | Fuel Cell |
| GHG | Green House Gas |
| H2 | Hydrogen |
| HP | Heat Pump |
| IEA | International Energy Agency |
| IPCC | Intergovernmental Panel on Climate Change |
| IRENA | International Renewable Energy Agency |
| ISPRA | Istituto Superiore per la Protezione e la Ricerca Ambientale |
| JRC | Joint Research Centre |
| LDV | Light Duty Vehicles |
| Li-ion | Lithium-ion |
| MOA | Multi-Objective Approach |
| MOO | Multi-Objective Optimization |
| NG | Natural Gas |
| OCGT | Open Cycle Gas Turbine |
| OECD | Organization for Economic Co-operation and Development |
| Oemof | Open Energy Modelling Framework |
| O&M | Operation and Maintenance |
| P2G | Power to Gas |
| PHS | Pumped Hydro Storage |
| PV | Photovoltaic |
| RES | Renewable Energy Source |
| TES | Thermal Energy Storage |
| TSO | Transmission System Operator |
| US | Utility Scale |
| wacc | Weighted Average Cost of Capital |

**Declaration of generative AI and AI-assisted technologies in the writing process**

During the preparation of this work the authors used ChatGPT to edit and rephrase some sentences. After using this tool/service, the authors reviewed and edited the content as needed and take full responsibility for the content of the publication.

**References**

[1] Sixth Assessment Report — IPCC n.d. https://www.ipcc.ch/assessment-report/ar6/ (accessed November 28, 2022).
[2] Environment UN. Emissions Gap Report 2020. UNEP - UN Environ Programme 2020. http://www.unep.org/emissions-gap-report-2020 (accessed November 28, 2022).
[3] Key aspects of the Paris Agreement | UNFCCC n.d. https://unfccc.int/most-requested/key-aspects-of-the-paris-agreement (accessed November 28, 2022).
[4] Ritchie H, Roser M, Rosado P. $CO_2$ and Greenhouse Gas Emissions. Our World Data 2020.
[5] Fit for 55 n.d. https://www.consilium.europa.eu/en/policies/green-deal/fit-for-55-the-eu-plan-for-a-green-transition/ (accessed November 18, 2022).
[6] PNIEC_finale_17012020.pdf n.d.

[7] Bella AD. oemof_Italy 2022.
[8] Mearns E, Sornette D. Are 2050 energy transition plans viable? A detailed analysis of projected Swiss electricity supply and demand in 2050. Energy Policy 2023;175:113347. https://doi.org/10.1016/j.enpol.2022.113347.
[9] Prina MG, Casalicchio V, Kaldemeyer C, Manzolini G, Moser D, Wanitschke A, et al. Multi-objective investment optimization for energy system models in high temporal and spatial resolution. Appl Energy 2020;264:114728. https://doi.org/10.1016/j.apenergy.2020.114728.
[10] Bellocchi S, Manno M, Noussan M, Prina MG, Vellini M. Electrification of transport and residential heating sectors in support of renewable penetration: Scenarios for the Italian energy system. Energy 2020;196:117062. https://doi.org/10.1016/j.energy.2020.117062.
[11] Hilpert S, Dettner F, Al-Salaymeh A. Analysis of Cost-Optimal Renewable Energy Expansion for the Near-Term Jordanian Electricity System. Sustainability 2020;12:9339. https://doi.org/10.3390/su12229339.
[12] Herc L, Pfeifer A, Feijoo F, Duić N. Energy system transitions pathways with the new H2RES model: A comparison with existing planning tool. E-Prime - Adv Electr Eng Electron Energy 2021;1:100024. https://doi.org/10.1016/j.prime.2021.100024.
[13] Louis J-N, Allard S, Kotrotsou F, Debusschere V. A multi-objective approach to the prospective development of the European power system by 2050. Energy 2020;191:116539. https://doi.org/10.1016/j.energy.2019.116539.
[14] Frysztacki MM, Hörsch J, Hagenmeyer V, Brown T. The strong effect of network resolution on electricity system models with high shares of wind and solar. Appl Energy 2021;291:116726. https://doi.org/10.1016/j.apenergy.2021.116726.
[15] Victoria M, Zhu K, Brown T, Andresen GB, Greiner M. The role of storage technologies throughout the decarbonisation of the sector-coupled European energy system. Energy Convers Manag 2019;201:111977. https://doi.org/10.1016/j.enconman.2019.111977.
[16] Hilpert S, Kaldemeyer C, Krien U, Günther S, Wingenbach C, Plessmann G. The Open Energy Modelling Framework (oemof) - A new approach to facilitate open science in energy system modelling. Energy Strategy Rev 2018;22:16–25. https://doi.org/10.1016/j.esr.2018.07.001.
[17] ALLEGATO A.24 AL CODICE DI RETE: INDIVIDUAZIONE ZONE DELLA RETE RILEVANTE - PDF Free Download n.d. https://docplayer.it/9456044-Allegato-a-24-al-codice-di-rete-individuazione-zone-della-rete-rilevante.html (accessed November 18, 2022).
[18] Piano di Sviluppo 2021 n.d.:372.
[19] 6-CONSUMI_8d9cfa23d9b95aa.pdf n.d.
[20] Le Centrali in Italia – Assocarboni n.d. https://assocarboni.it/assocarboni/il-carbone/le-centrali-in-italia/ (accessed November 18, 2022).
[21] JRC Hydro-power database - Data Europa EU n.d. https://data.europa.eu/data/datasets/52b00441-d3e0-44e0-8281-fda86a63546d?locale=en (accessed November 28, 2022).
[22] Fonti rinnovabili - Terna spa n.d. https://www.terna.it/it/sistema-elettrico/dispacciamento/fonti-rinnovabili (accessed November 28, 2022).
[23] relazione_annuale_situazione_energetica_nazionale_dati_2019.pdf n.d.
[24] 3-IMPIANTI DI GENERAZIONE_8d9cf9484bbf3bd.pdf n.d.
[25] Mongird K, Viswanathan V, Alam J, Vartanian C, Sprenkle V, Baxter R. 2020 Grid Energy Storage Technology Cost and Performance Assessment 2020:117.
[26] Energy System Technology Data 2022.
[27] Coal Data Browser - Aggregate coal mine production for all coal n.d. https://www.eia.gov/coal/data/browser/ (accessed November 28, 2022).

[28] Natural gas price statistics n.d. https://ec.europa.eu/eurostat/statistics-explained/index.php?title=Natural_gas_price_statistics (accessed November 28, 2022).
[29] Pubblicazioni Statistiche - Terna spa n.d. https://www.terna.it/it/sistema-elettrico/statistiche/pubblicazioni-statistiche (accessed November 18, 2022).
[30] Physical Foreign flow - Terna spa n.d. https://www.terna.it/en/electric-system/transparency-report/physical-foreign-flow (accessed January 12, 2023).
[31] Download center - Terna spa n.d. https://www.terna.it/en/electric-system/transparency-report/download-center (accessed November 18, 2022).
[32] 5-PRODUZIONE_8d9cf942c92ff41.pdf n.d.
[33] Italy - Countries & Regions - IEA n.d. https://www.iea.org/countries/italy (accessed November 28, 2022).
[34] Italy - OECD n.d. https://www.oecd.org/italy/ (accessed November 28, 2022).
[35] bp-stats-review-2019-full-report.pdf n.d.
[36] Italy_Europe_RE_SP.pdf n.d.
[37] European Commission. Joint Research Centre. GHG emissions of all world: 2021 report. LU: Publications Office; 2021.
[38] Rapporto_220_2015.pdf n.d.
[39] 201001_ITALY HYDROPOWER.pdf n.d.
[40] Manzella A, Serra D, Cesari G, Bargiacchi E, Cei M, Conti P, et al. Geothermal Energy Use, Country Update for Italy n.d.
[41] Econnextion: la mappa delle connessioni rinnovabili - Terna spa n.d. https://www.terna.it/it/sistema-elettrico/rete/econnextion (accessed March 3, 2023).
[42] Tröndle T, Pfenninger S, Lilliestam J. Home-made or imported: On the possibility for renewable electricity autarky on all scales in Europe. Energy Strategy Rev 2019;26:100388. https://doi.org/10.1016/j.esr.2019.100388.
[43] Lund H, Thellufsen JZ, Østergaard PA, Sorknæs P, Skov IR, Mathiesen BV. EnergyPLAN – Advanced analysis of smart energy systems. Smart Energy 2021;1:100007. https://doi.org/10.1016/j.segy.2021.100007.
[44] Fact sheets about Photovoltaics - ETIP PV n.d. https://etip-pv.eu/publications/fact-sheets/ (accessed November 28, 2022).
[45] Tsiropoulos I, Tarvydas D, Zucker A. Cost development of low carbon energy technologies: Scenario-based cost trajectories to 2050, 2017 edition. JRC Publ Repos 2018. https://doi.org/10.2760/490059.
[46] Vartiainen E, Masson G, Breyer C, Moser D, Román Medina E. Impact of weighted average cost of capital, capital expenditure, and other parameters on future utility-scale PV levelised cost of electricity. Prog Photovolt Res Appl 2020;28:439–53. https://doi.org/10.1002/pip.3189.
[47] Chen T, Jin Y, Lv H, Yang A, Liu M, Chen B, et al. Applications of Lithium-Ion Batteries in Grid-Scale Energy Storage Systems. Trans Tianjin Univ 2020;26:208–17. https://doi.org/10.1007/s12209-020-00236-w.
[48] Cole W, Frazier AW, Augustine C. Cost Projections for Utility-Scale Battery Storage: 2021 Update. Renew Energy 2021:21.
[49] Steward DM. Scenario Development and Analysis of Hydrogen as a Large-Scale Energy Storage Medium n.d.:31.
[50] Batas Bjelić I, Rajaković N. Simulation-based optimization of sustainable national energy systems. Energy 2015;91:1087–98. https://doi.org/10.1016/j.energy.2015.09.006.
[51] European Green Deal - Delivering on our targets. Eur Comm - Eur Comm n.d. https://ec.europa.eu/commission/presscorner/detail/en/fs_21_3688 (accessed November 28, 2022).

[52] Energy transition and carbon budget: the Italian scenario. EURAC Res n.d. https://www.eurac.edu/en/institutes-centers/institute-for-renewable-energy/tools-services/energy-modelling/energy-transition-and-carbon-budget-the-italian-scenario (accessed January 12, 2023).
[53] Italy. National Communication (NC). NC 8. | UNFCCC n.d. https://unfccc.int/documents/624766 (accessed January 16, 2023).
[54] content/enel-com/en/authors/simone-mori. 'Fit for 55': EU on track for climate goals and sustainable growth n.d. https://www.enel.com/company/stories/articles/2021/08/fit-for-55-europe-energy-transition-goals (accessed November 19, 2022).
[55] Solar PV – Analysis. IEA n.d. https://www.iea.org/reports/solar-pv (accessed January 17, 2023).
[56] Kharel S, Shabani B. Hydrogen as a Long-Term Large-Scale Energy Storage Solution to Support Renewables. Energies 2018;11:2825. https://doi.org/10.3390/en11102825.
[57] Egeland-Eriksen T, Hajizadeh A, Sartori S. Hydrogen-based systems for integration of renewable energy in power systems: Achievements and perspectives. Int J Hydrog Energy 2021;46:31963–83. https://doi.org/10.1016/j.ijhydene.2021.06.218.
[58] Colbertaldo P, Guandalini G, Campanari S. Modelling the integrated power and transport energy system: The role of power-to-gas and hydrogen in long-term scenarios for Italy. Energy 2018;154:592–601. https://doi.org/10.1016/j.energy.2018.04.089.
[59] Creutzig F, Niamir L, Bai X, Callaghan M, Cullen J, Díaz-José J, et al. Demand-side solutions to climate change mitigation consistent with high levels of well-being. Nat Clim Change 2022;12:36–46. https://doi.org/10.1038/s41558-021-01219-y.
[60] Di Bella A, Tavoni M. Demand-side policies for power generation in response to the energy crisis: a model analysis for Italy 2022. https://doi.org/10.48550/arXiv.2212.06744.
[61] Mundaca L, Ürge-Vorsatz D, Wilson C. Demand-side approaches for limiting global warming to 1.5 °C. Energy Effic 2019;12:343–62. https://doi.org/10.1007/s12053-018-9722-9.
[62] CPUC Issues Independent Study on Injecting Hydrogen Into Natural Gas Systems n.d. https://www.cpuc.ca.gov/news-and-updates/all-news/cpuc-issues-independent-study-on-injecting-hydrogen-into-natural-gas-systems (accessed January 17, 2023).
[63] Melaina MW, Antonia O, Penev M. Blending Hydrogen into Natural Gas Pipeline Networks: A Review of Key Issues. Renew Energy 2013.
[64] Hoekstra A, Steinbuch M, Verbong G. Creating Agent-Based Energy Transition Management Models That Can Uncover Profitable Pathways to Climate Change Mitigation. Complexity 2017;2017:1–23. https://doi.org/10.1155/2017/1967645.
[65] Kogler G. Impacts of Government Incentives on Solar Power Growth - GK-Electrics 2022. https://gk-electrics.com/en/government-incentives/ (accessed September 6, 2023).
[66] Italy Climate Resilience Policy Indicator – Analysis. IEA n.d. https://www.iea.org/articles/italy-climate-resilience-policy-indicator (accessed September 8, 2023).
[67] Bartos MD, Chester MV. Impacts of climate change on electric power supply in the Western United States. Nat Clim Change 2015;5:748–52.
[68] Tobin I, Greuell W, Jerez S, Ludwig F, Vautard R, Vliet MTH van, et al. Vulnerabilities and resilience of European power generation to 1.5 °C, 2 °C and 3 °C warming. Environ Res Lett 2018;13:044024. https://doi.org/10.1088/1748-9326/aab211.
[69] He G, Mallapragada DS, Bose A, Heuberger-Austin CF, Gençer E. Sector coupling via hydrogen to lower the cost of energy system decarbonization. Energy Environ Sci 2021;14:4635–46. https://doi.org/10.1039/D1EE00627D.
[70] Gea-Bermúdez J, Jensen IG, Münster M, Koivisto M, Kirkerud JG, Chen Y, et al. The role of sector coupling in the green transition: A least-cost energy system development in

Northern-central Europe towards 2050. Appl Energy 2021;289:116685. https://doi.org/10.1016/j.apenergy.2021.116685.
[71] Prina MG, Nastasi B, Groppi D, Misconel S, Garcia DA, Sparber W. Comparison methods of energy system frameworks, models and scenario results. Renew Sustain Energy Rev 2022;167:112719. https://doi.org/10.1016/j.rser.2022.112719.
[72] Pfenninger S. Dealing with multiple decades of hourly wind and PV time series in energy models: A comparison of methods to reduce time resolution and the planning implications of inter-annual variability. Appl Energy 2017;197:1–13. https://doi.org/10.1016/j.apenergy.2017.03.051.
[73] Casalicchio V, Manzolini G, Prina MG, Moser D. From investment optimization to fair benefit distribution in renewable energy community modelling. Appl Energy 2022;310:118447. https://doi.org/10.1016/j.apenergy.2021.118447.
[74] Bistline JET. The importance of temporal resolution in modeling deep decarbonization of the electric power sector. Environ Res Lett 2021;16:084005. https://doi.org/10.1088/1748-9326/ac10df.
[75] Poncelet K, Delarue E, Six D, Duerinck J, D'haeseleer W. Impact of the level of temporal and operational detail in energy-system planning models. Appl Energy 2016;162:631–43. https://doi.org/10.1016/j.apenergy.2015.10.100.
[76] Kotzur L, Markewitz P, Robinius M, Stolten D. Time series aggregation for energy system design: Modeling seasonal storage. Appl Energy 2018;213:123–35. https://doi.org/10.1016/j.apenergy.2018.01.023.
[77] Zurita A, Mata-Torres C, Cardemil JM, Escobar RA. Assessment of time resolution impact on the modeling of a hybrid CSP-PV plant: A case of study in Chile. Sol Energy 2020;202:553–70. https://doi.org/10.1016/j.solener.2020.03.100.
[78] Frew BA, Jacobson MZ. Temporal and spatial tradeoffs in power system modeling with assumptions about storage: An application of the POWER model. Energy 2016;117:198–213. https://doi.org/10.1016/j.energy.2016.10.074.